\documentclass[onecolumn,aps,showpacs,preprintnumbers,amssymb,superscriptaddress,nofootinbib,groupedaddress]{revtex4-2}

\usepackage{graphicx,amsmath,booktabs,appendix}
\usepackage[usenames, dvipsnames]{color}
\usepackage{lineno}

\newcommand{\beq}{\begin{equation}}
\newcommand{\eeq}{\end{equation}}
\newcommand{\bea}{\begin{eqnarray}}
\newcommand{\eea}{\end{eqnarray}}

\begin{document}

\title{Longitudinal single-bunch instabilities driven by coherent undulator radiation in the cavity modulator of a steady-state microbunching storage ring}
\author{Cheng-Ying Tsai}\email[]{\texttt{jcytsai@hust.edu.cn}}
\affiliation{School of Electrical and Electronic Engineering, Huazhong University of Science and Technology, Wuhan, 430074, China}

\date{\today}


\begin{abstract}
Recently a mechanism of storage ring operation based on the steady-state microbunching has been proposed and investigated, which contains a laser cavity modulator providing the longitudinal focusing for the circulating microbunches. In this paper we analyze the impact of the coherent undulator radiation on the longitudinal single-bunch multi-turn collective dynamics, exploring a new {possible} instability mechanism. We formulate the multi-turn equations of motion for the single microbunch as two sets of difference equations in the modulator and in the remaining storage ring. The dispersion equation can then be obtained by introduction of the undulator-averaged phase space coordinates. The predicted instability growth rate shows reasonable agreement with the numerical turn-by-turn tracking simulations provided validity of the underlying assumptions. The analysis shall provide some insights for the coherent undulator radiation driven multi-turn instability in the cavity modulator. Differences between such a new instability mechanism and the Robinson instability in a conventional radio-frequency-based storage ring are also discussed.
\end{abstract}

\maketitle

\section{Introduction}\label{SecI}
Accelerator-based light sources have played a significant role in various fields of science for the past three decades. Nowadays the advanced light sources based on particle accelerators may be classified into three categories: the free electron lasers (FEL) driven by linear accelerators (linac), the synchrotron light sources based on the storage ring accelerators, or those based on energy recovery linacs (ERL). The linac-based single-pass FELs have been known to provide the peak brightness six to eight orders of magnitude higher than the storage-ring-based synchrotron light sources, thanks to the microbunching process in FEL instability~\cite{Ref01}. The relatively low average brightness of linac FEL, attributed to the low repetition rate of the linear accelerators, can be improved by the superconducting radio-frequency (RF) technology, e.g., the ERL-based fourth generation light sources. The peak brightness is the primary concern along this line of thought. 

The storage-ring-based synchrotron light sources have an excellent average brightness due to high repetition rate and high revolution frequencies. The synchrotron radiation spans a continuous, wide spectrum covering from infrared, through visible, down to x-ray wavelength ranges. However, the circulating electron beam of about millimeter (mm) bunch length is usually not microbunched, thus only producing incoherent synchrotron radiation (ISR). The peak power from storage ring light sources is much lower than that from the linac-based FEL. Recently a mechanism of the steady-state microbunching (SSMB) in a storage ring has been proposed and investigated~\cite{Ref02,Ref03,Ref04,Ref05,Ref06,Ref07,Ref08}. The SSMB aims to maintain the same excellent high repetition rate, close to continuous-wave operation, as the storage ring. Moreover, replacing the conventional RF cavity with a laser modulator for longitudinal focusing, the individual electron bunches can be microbunched in a steady state. The microbunched electron bunch train, with individual bunch length comparable to or shorter than the radiation wavelength, can thus produce coherent and powerful synchrotron radiations. Pursuing this further, the name of the game is the average brightness, in contrast to the aforementioned linac-based FELs.

{It has been known that in a conventional storage ring a typical RF cavity plays two roles: compensate for the beam energy loss (“acceleration”), and provide the circulating beam with the longitudinal focusing (“bunching” or “modulation”). In contrast, in the design of an SSMB storage ring, the laser modulator cavity is assumed to play the only role of bunching/modulating the microbunched beam. The compensation for the energy loss of a microbunched beam will be considered by introduction of an induction linac. The induction pulse may cover part of a revolution every revolution time sufficient to cover the filled microbunch train (which contains many microbunches with bunch spacing equal to the laser modulation wavelength). }

To ensure high coherent radiation power output, the SSMB-based light sources rely on the two enhancement mechanisms: the individual microbunches and the microbunch train. The former requires that the individual bunch length be comparable to or shorter than the radiation wavelength; the latter suggests a sufficient number of microbunches. Here we notice that, whether the linac-based or storage-ring-based light sources, their output performance enhancement involves improvement of the electron beam brightness. A high-brightness electron beam is characteristic of small emittances, low beam energy spread and even a short bunch length, with as many as possible electrons in a beam phase space volume. The high-brightness charged particle beam is inevitably accompanied by the collective instabilities~\cite{Ref09,Ref10}.

The theoretical formulation of microbunching instability has been developed, for example, in single-pass linac-based FEL facilities~\cite{Ref11,Ref12,Ref13,Ref14,Ref15,Ref16,Ref17,Ref18}, the RF-based storage ring light sources~\cite{Ref19,Ref20,Ref21,Ref22,Ref23} and the ERL-based light sources or recirculating electron cooling facilities~\cite{Ref24,Ref25,Ref26,Ref27,Ref28}. As for the SSMB-based storage ring light sources, the single-particle dynamics has been recently investigated in detail~\cite{Ref06,Ref29} and the proof-of-principle experiment has been conducted to demonstrate that the microbunching is achievable after one complete revolution of an electron bunch modulated by an external laser in a dedicated storage ring~\cite{Ref07}. However the collective effects in the SSMB storage ring were not examined in detail. As the beam intensity or bunch charge increases, the collective effects emerge. Due to replacement of an RF cavity by the laser modulator, the bunch spacing of an SSMB-like microbunch train can be five orders of magnitude shorter than that in conventional RF-based storage rings. In the situation when the undulator configuration still remains the same, the radiation wake function generated by the traversing electron bunch can now introduce a relatively long-range collective effect, albeit it has been viewed as a short-range single-bunch effect in the conventional storage rings. Such an interesting \textit{long-range} situation may bring about new instability mechanisms, e.g., the single-pass multi-bunch instability~\cite{Ref08} or the multi-turn single-bunch/multi-bunch instability.

Recently we have developed a theoretical formulation for the single-pass multi-bunch longitudinal beam breakup (BBU) instability driven by the coherent undulator radiation~\cite{Ref08}. The studies indicate that such BBU instability in principle would not affect the SSMB performance according to the preliminary design parameters~\cite{Ref04}. However, for an advanced design of an SSMB storage ring, the requirement of a stronger laser modulation to provide a stronger longitudinal focusing may need to form the laser modulator as an enhancement cavity~\cite{Ref30,Ref31}. In this laser modulator cavity the undulator radiation wake may have another long-range, multi-turn, single-bunch collective effect, as we will explore in this paper. Here we give a qualitative description about this potential long-range collective instability as follows. A laser modulator cavity is comprised of an external laser, a set of cavity mirrors, and an undulator along which the laser and the electron resonate and co-propagate. The external laser and the undulator magnetic field will form the phase space buckets, similar to the role of an RF cavity in a storage ring~\cite{Ref33,Ref35}. When an electron bunch traverses the modulator undulator, it emits coherent undulator radiation~\cite{Ref31a,Ref31b}. After the electron bunch leaves the laser modulator cavity, it would no longer emit radiations but the generated radiation fields would be confined inside the cavity because the undulator is by design tuned to resonate the electron bunch with the external laser. Due to the resonance condition the undulator radiation wavelength is close to the central wavelength of the external laser and can be stored inside the cavity for a while. To ensure effective interaction of the external laser with the circulating electron bunch (in order to provide longitudinal focusing), their relative phase, or the timing, shall be locked. Such a phase lock guarantees that the electron bunch would meet the same phase space bucket, when completing a revolution and returning back to the undulator entrance. To maintain such a phase lock between the external laser and the circulating microbunch is indeed a technically challenging issue on a turn-by-turn basis. Inside the undulator, the electron bunch would receive consecutive energy kicks of radiation wake functions from the previous turns. One might expect that, if the radiation fields being stored in the modulator cavity stay for a long time, a circulating electron beam perturbed by the radiation fields turn by turn may eventually become unstable. It turns out that this instability mechanism is similar to but not the same as Robinson instability, as occurred in the conventional RF-based storage rings~\cite{Ref09,Ref10}. {It deserves here to remark that the radiation induced in the modulator undulator of the laser cavity is not the radiation intended by the SSMB concept. It is the downstream radiator undulator that emits the radiation by the SSMB design. }

In this paper we study the longitudinal single-bunch instabilities driven by the coherent undulator radiation in the cavity modulator of an SSMB storage ring. We will begin from constructing the undulator radiation wake function and formulate the macroparticle equations of motions in the SSMB dynamical system. Then we solve for the system (in)stability and compare our theoretical predictions with turn-by-turn tracking simulations. A few differences between the coherent radiation induced multi-turn instability and the conventional RF cavity induced long-range collective instabilities may be summarized in the following two aspects. First, unlike the conventional wake function, the radiation wake is not localized at a specific $s$ position, where $s$ is the global path-length coordinate. In our case the same bunch at later times, or the following microbunches, do not only sample the temporally decaying wake field. Instead, the radiation fields propagate, bounce back and forth among the cavity mirrors, and meet/overlap the electron bunch completing one revolution in the remaining storage ring. Moreover, the radiation fields are with finite duration because the undulator length is finite.

It deserves here to emphasize the contributions of this work. First, we formulate and analyze the coherent undulator radiation induced single-bunch multi-turn instability. Second, we apply the analysis to the recently proposed SSMB storage ring and explore the instability behavior based on the preliminary design parameters. Third, we make a brief comparison between such an instability and the classic Robinson instability. {It is also worthwhile to distinguish the free electron laser oscillator (FELO) from the SSMB concept and highlight the subject of this work in the following aspects. For FELO, the cavity oscillator is the key component that produces the output (the desired radiation). Initially there is no radiation field inside the cavity. The radiation fields are built up from shot noise of the usually unbunched electron beam. For SSMB, the radiator undulator (not modulator undulator) is the key component that produces the output, desired radiation. Although the radiator undulator is not the focus in this work, in the present preliminary design it is a single-pass device and does not form an oscillator. The radiation fields are built up from single-pass coherent radiation from a single microbunch (or microbunch train). For FELO, the electron beam traverses the oscillator once or a few times. When the beam quality degrades, a new beam bunch may replace. For storage-ring based FELO, one may wait for a few radiation damping times to restore the beam quality. For SSMB, the radiation fields are produced in the single-pass radiator. There is no explicit multi-turn feedback mechanism between the circulating microbunch and the emitted radiation fields in the radiator. In the zeroth-order analysis, the laser modulator cavity acts as a buncher (or modulator). The radiation emitted in the modulator undulator is ignored in the single-particle dynamics (or, pure optics). Based on the zeroth-order studies, in this work we perform the first-order analysis by considering the radiation fields in the laser modulator cavity. Different from FELO, in the SSMB laser modulator cavity the external laser field is injected and serves to longitudinally bunch or modulate the circulating microbunch. The intention is not to store the emitted radiation fields. For FELO, a typical analysis usually assumes a coasting beam, where the bunch length is much larger than the radiation wavelength. In this work, in the laser modulator cavity the bunch length is much smaller than the modulation laser wavelength (also the radiation wavelength in the modulator). } Here we remind that a more in-depth and quantitative discussion, including possible connection to the high-gain FEL instability~\cite{Ref32}, will be published in a separate paper elsewhere.

This paper is organized as follows. In Sec.~\ref{SecIIA} we summarize the results for calculation of the undulator impedance and the radiation wake function. Then in Sec.~\ref{SecIIB} we deduce the argument of the wake function by examining the laser-electron-radiation interaction in the undulator and the remaining storage ring. Using time-domain description, in Sec.~\ref{SecIIC} we formulate the macroparticle equations of motion and would obtain two sets of coupled difference equations. In absence of radiation fields, the equations of motion are reduced to the pure optics case. Some basic results are discussed in Sec.~\ref{SecIID} for such a trivial case. In the presence of radiation fields, the difference equations are then solved in Sec.~\ref{SecIII} using z-transform to obtain the dispersion relation and determine the system (in)stability. Having developed the theoretical formulation, we apply it for a systematic, quantitative analysis of the preliminary SSMB storage ring design parameters~\cite{Ref04} in Sec.~\ref{SecIV}. In Sec.~\ref{SecV} we briefly compare the new instability mechanism with Robinson instability and outline possible connection to the FEL instability. Finally we summarize this work in Sec.~\ref{SecVI}.

\section{Theoretical formulation}\label{SecII}
\begin{figure}
\centering
\includegraphics[width=5in]{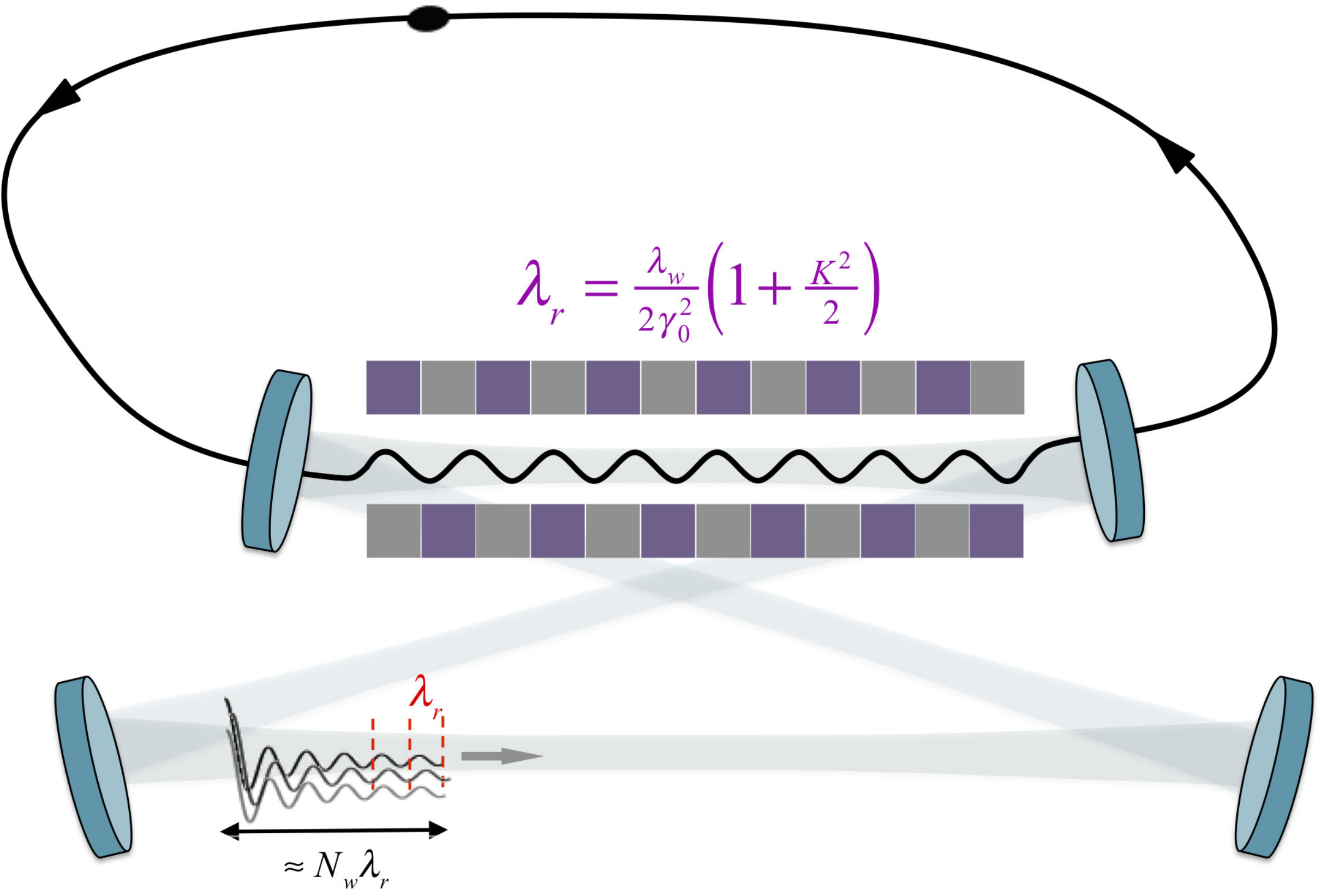}
\caption{\textmd{Schematic layout of a storage ring with a laser ring-cavity modulator. Inside the cavity the laser fields, together with the undulator magnetic field, provide longitudinal focusing to the circulating electron bunch. Upon each passage through the $N_w$-period undulator, an electron will produce radiation wake fields, each with the finite duration $N_w \lambda_r$ with $\lambda_r$ the undulator resonant wavelength. The propagating radiation fields bounce back and forth among the mirrors in the cavity modulator.}}
\label{Fig1}
\end{figure}

Figure~\ref{Fig1} shows the schematic layout of a storage ring with a laser modulator cavity. Let us assume an electron bunch circulates in the storage ring. Similar to the role of a RF cavity in the conventional storage ring, the laser modulator provides longitudinal focusing to the beam. The modulator consists of an external laser, a set of cavity mirrors to accumulate a sufficient laser intensity, and an undulator. To provide effective longitudinal focusing, a resonant interaction between the laser and the electron bunch must be ensured, i.e., the laser and the electron beam have to satisfy the resonance condition. Upon each passage through the $N_w$-period undulator, an electron will produce radiation fields, each with the finite duration $N_w \lambda_r$, with $\lambda_r$ the undulator resonant wavelength. The propagating radiation fields bounce back and forth among the mirrors in the modulator and must meet the circulating electron bunch in a synchronous way.

Below we shall first summarize the results for the undulator radiation impedance and the corresponding wake function. We then construct the relation of the relative positions between the radiation wake and the electron bunch along the undulator between arbitrary two passages. For simplicity a microbunch is represented as a macroparticle with a total bunch charge $Ne$.

\subsection{Undulator radiation impedance and wake function}\label{SecIIA}
When an electron traverses an undulator, it emits undulator radiation. The radiation wake function $W_{\parallel}$ can be obtained by inverse Fourier transformation of the radiation impedance~\cite{Ref09}
\begin{equation}\label{Eq1}
{W_\parallel }(z > 0) = \frac{{2c}}{\pi }\int_0^\infty  {\operatorname{Re} {Z_\parallel }(k)\cos kz{\text{d}}k}, 
\end{equation}
where $c$ is the speed of light in vacuum, $k$ is the wavenumber, and Re takes the real part of the radiation impedance $Z_{\parallel}$. Here $z > 0$ refers to the bunch head and $z < 0$ to the bunch tail. In this expression we require ${W_\parallel }(z < 0) = 0$. In what follows we may need to evaluate the derivative of the wake function with respect to longitudinal bunch coordinate $z$
\begin{equation}\label{Eq2}
{W_\parallel^{\prime} }(z > 0) =  - \frac{{2c}}{\pi }\int_0^\infty  {k\operatorname{Re} {Z_\parallel }(k)\sin kz{\text{d}}k}.
\end{equation}

The undulator radiation impedance per unit length per unit undulator wavenumber at a specific harmonic can be written as~\cite{Ref33,Ref34,Ref35}
\begin{equation}\label{Eq3}
\frac{{\operatorname{Re} {Z_{\parallel ,H}}(\omega )}}{{{L_w}{k_w}}} = {Z_0}\bar G({\theta _1})\left( {1 + \tfrac{{{K^2}}}{2}} \right){\left( {\frac{\omega }{{{\omega _r}}}} \right)^{ - 1}},
\end{equation}
where $Z_0 \approx 377$ $\Omega$ is the free space impedance, $k_w = 2\pi/\lambda_w$ with $\lambda_w$ the undulator period, $L_w = N_w \lambda_w$ the undulator length with $N_w$ the number of undulator periods. In Eq. (\ref{Eq3}), the frequency-dependent polar angle is ${\theta _1} = \tfrac{1}{{{\gamma _0}}}\sqrt {\left( {1 + \tfrac{{{K^2}}}{2}} \right)\left( {H\tfrac{{{\omega _r}}}{\omega } - 1} \right)}$, where the positive integer $H$ is the harmonic number, $K$ is dimensionless undulator parameter, $\gamma_0$ is the electron reference energy, and the resonant frequency ${\omega _r} = c{k_r} = {{2\pi c} \mathord{\left/ {\vphantom {{2\pi c} {{\lambda _r}}}} \right. \kern-\nulldelimiterspace} {{\lambda _r}}}$ with the undulator resonant wavelength $\lambda_r$
\begin{equation}\label{Eq4}
{\lambda _r} = \frac{{{\lambda _w}}}{{2\gamma _0^2}}\left( {1 + \frac{{{K^2}}}{2}} \right).
\end{equation}

Moreover, the azimuthal angle $\phi$-averaged function
\begin{equation}\label{Eq5}
\bar G(\theta ) = \frac{1}{{2\pi }}\int_0^{2\pi } {G(\theta ,\phi ){\text{d}}\phi } 
\end{equation}
with $G(\theta ,\phi ) = {G_\sigma }(\theta ,\phi ) + {G_\pi }(\theta ,\phi )$
\begin{equation}\label{Eq6}
{G_\sigma }(\theta ,\phi ) = \frac{1}{2}{\left[ {\frac{{H\left( {K{{\mathcal{D}}_1} + {\gamma _0}\theta {{\mathcal{D}}_2}\cos \phi } \right)}}{{1 + \tfrac{{{K^2}}}{2} + \gamma _0^2{\theta ^2}}}} \right]^2}
\end{equation}
and
\begin{equation}\label{Eq7}
{G_\pi }(\theta ,\phi ) = \frac{1}{2}{\left( {\frac{{H{\gamma _0}\theta {{\mathcal{D}}_2}\sin \phi }}{{1 + \tfrac{{{K^2}}}{2} + \gamma _0^2{\theta ^2}}}} \right)^2}
\end{equation}
where
\begin{equation}\label{Eq8}
{{\mathcal{D}}_1} =  - \frac{1}{2}\sum\limits_{m =  - \infty }^\infty  {{J_{H + 2m - 1}}\left( {H\alpha } \right)\left[ {{J_m}\left( {H\zeta } \right) + {J_{m - 1}}\left( {H\zeta } \right)} \right]} 
\end{equation}
and
\begin{equation}\label{Eq9}
{{\mathcal{D}}_2} = \sum\limits_{m =  - \infty }^\infty  {{J_{H + 2m}}\left( {H\alpha } \right){J_m}\left( {H\zeta } \right)} 
\end{equation}
with $\alpha  = \frac{{2K{\gamma _0}\theta \cos \phi }}{{1 + \tfrac{{{K^2}}}{2} + \gamma _0^2{\theta ^2}}},\zeta  = \frac{{{{{K^2}} \mathord{\left/
 {\vphantom {{{K^2}} 4}} \right.
 \kern-\nulldelimiterspace} 4}}}{{1 + \tfrac{{{K^2}}}{2} + \gamma _0^2{\theta ^2}}}$.

When multiple harmonics are included, the overall radiation impedance function $\text{Re}Z_{\parallel}$ can be obtained by summing the individual harmonics $H$ in Eq. (\ref{Eq3}), i.e., $\operatorname{Re} {Z_\parallel }(\omega ) = \sum\limits_H {\operatorname{Re} {Z_{\parallel ,H}}(\omega )}$. The corresponding radiation wake function can be calculated via Eq. (\ref{Eq1}). A proper choice of the maximum harmonic number $H_{\text{max}}$ in the undulator radiation spectrum depends on the design goal of the radiator and the uncorrelated beam energy spread~\cite{Ref08}. The equivalent momentum compaction factor for an undulator can be obtained from Eq. (\ref{Eq4}), giving $|R_{56}| = 2 N_w \lambda_r$. When traversing the undulator, the uncorrelated beam energy spread will induce longitudinal smearing $\delta z \approx |R_{56}|\sigma_{\delta}$. This smearing should be within the resonant wavelength of the radiator, denoted as $\lambda_r/H_{\text{max}}$. Thus the maximum harmonic number of our interest can be determined by $H_{\text{max}} \le 1/2N_w\sigma_{\delta}$. We notice that the macroparticle model adopted here can only reflect the motion of the bunch centroid or dipole motion. The effect of a finite beam energy spread has been excluded. Thus $H_{\text{max}}$ here is merely a free parameter.

\begin{figure}
\centering
\includegraphics[width=3.5in]{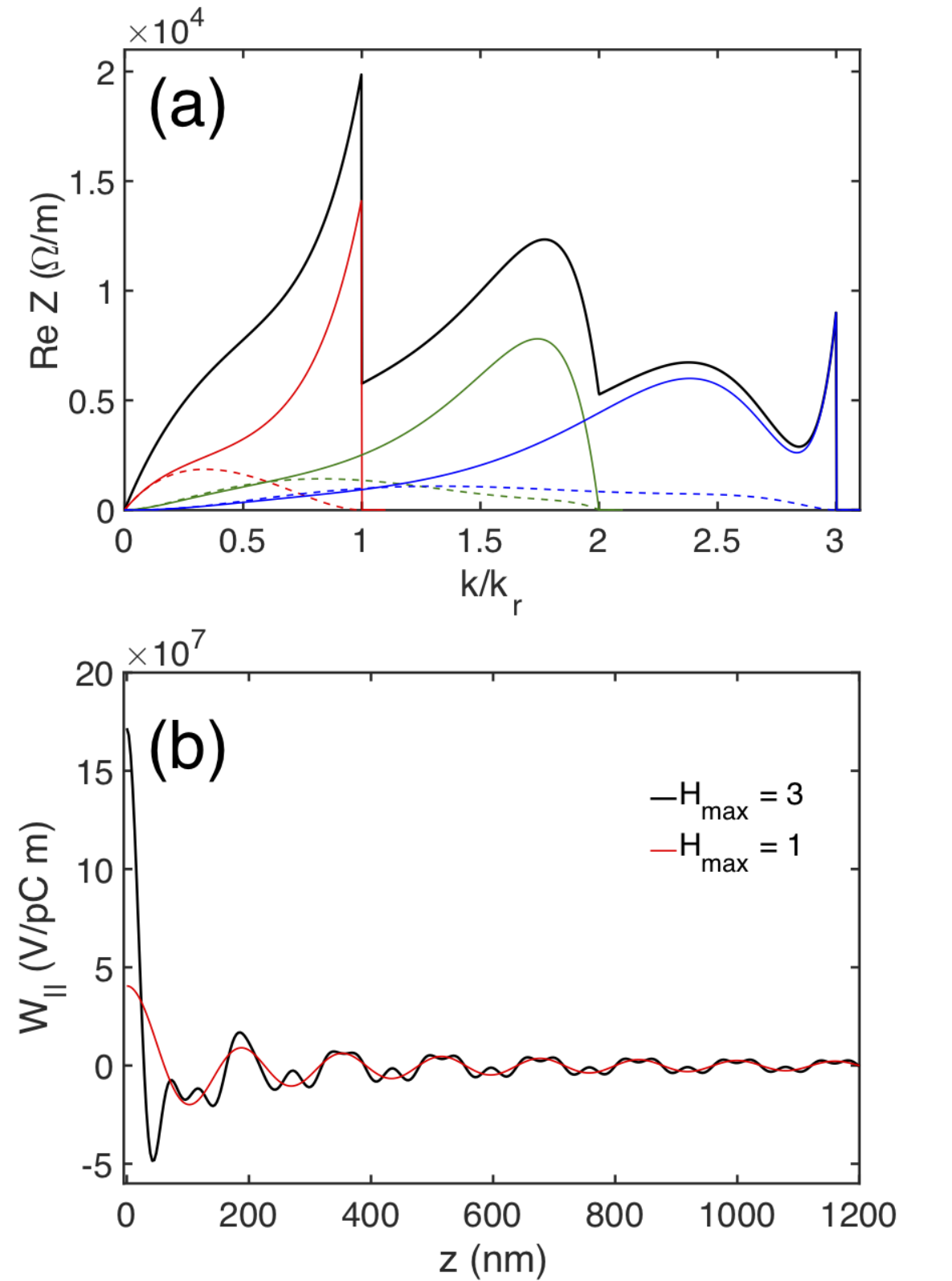}
\caption{\textmd{(a) Undulator radiation impedance per unit length and (b) the corresponding radiation wake function per unit length. The relevant parameters are $K = 4.2, \lambda_w =2$ cm, $\gamma_0 = \frac{400}{0.511} \approx 783$, $\lambda_r \approx 0.16$ $\mu$m. In (a), the red, green and blue lines are for the fundamental, second and third harmonics; with the dominant solid lines referring to the $\sigma$ mode and the dashed lines to the $\pi$ mode. The harmonic contents can be seen in (b) when including all three harmonics.}}
\label{Fig2}
\end{figure}

For the case of $H_{\text{max}} = 3$, the radiation impedance spectrum is shown in Fig.~\ref{Fig2}(a) and the corresponding wake function in Fig.~\ref{Fig2}(b) can be obtained by Eq. (\ref{Eq1}). It can be seen from Fig.~\ref{Fig2}(a) that the dominant contributions occur at $ k \approx k_r$ and their harmonics.

We also find that at each harmonic the spectrum tends to span over a wide, lower frequencies. This can be explained by the fact that the electron in the beam rest frame executes a figure-8 oscillation. Such figure-8 motion can be decomposed into an oscillation along the transverse horizontal $x$ axis with odd harmonics and the other along the longitudinal axis $z$ with even harmonics. In the electron rest frame, the dominant dipole oscillations along $x$ and $z$ will lead to dipole radiations emitted mainly perpendicular to the respective oscillation directions, i.e., in the longitudinal and transverse direction, respectively. After Lorentz transformation back to the laboratory frame, the off-axis Doppler red shift results in spectrum broadening at each harmonic, and moreover both the dipole radiations mainly lean in the forward $z$ direction. However, only the odd harmonics concentrate on $z$ axis. It is the two features that lead to the sharp, one-sided peaks around the odd harmonics. In addition, each harmonic contains two spectral components: the horizontal $\sigma$ mode [Eq. (\ref{Eq6})] and the vertical $\pi$ mode [Eq. (\ref{Eq7})]. Considering the planar undulator configuration, the dominant $\sigma$ mode has always a maximum on the axis. Figure 2(b) also shows the features of harmonics of the radiation wake (the $H_{\text{max}} = 3$ case).

As mentioned above, after the electron leaves from the undulator, the propagating radiation fields bounce back and forth among the mirrors in the modulator and may meet the circulating electron bunch at later times. Because the on-axis, resonant frequency of the undulator radiation is close to that of the external laser, the effects of the cavity mirrors on the radiation field may play a role and can be formulated as a field decay and frequency filtering~\cite{Ref01}. The cavity mirrors may reduce the field amplitude by a real multiplicative factor $\sqrt{R} = \sqrt{1-\alpha}$ with $\alpha$ the power loss and have a frequency-dependent effect
\begin{equation}\label{Eq10}
Z_\parallel ^c(\omega ) \approx \left( {1 - \frac{\alpha }{2}} \right){e^{ - \frac{{{\omega ^2}}}{{4\sigma _{{\text{refl}}}^2}}}}{Z_\parallel }(\omega ),
\end{equation}
where $\alpha \ll 1$ and a Gaussian filter with rms power bandwidth $\sigma_{\text{refl}}$ has been used. The ideal case refers to $\alpha = 0$ and $\sigma_{\text{refl}} \to \infty$.

In this subsection we have summarized how to obtain the undulator radiation wake function. In the next subsection we will deduce the argument of the wake function by examining the laser-electron-radiation interaction in the undulator and the remaining storage ring.

\subsection{Wake received by a circulating microbunch}\label{SecIIB}
This subsection is to determine the relative position between the generated radiation wakes and the electron bunch along the undulator between arbitrary two revolutions. To do so, we outline the two important situations. First, when an electron bunch traverses along the undulator and circulates in the storage ring, the effective interaction between the moving bunch and the external laser to provide longitudinal focusing is assured by the condition that they are synchronous or phase locked. That is, on each turn when the electron bunch arrives at the undulator entrance, it should always see the same phase space bucket, formed by the external laser and the undulator magnetic field. Second, an electron traverses along the undulator at a certain passage at the speed of $\beta_z c$, while both the emitted radiation and the external laser propagate at the speed of light $c$. Such speed difference leads to the slippage effect in the undulator; the radiation fields slip over one resonant wavelength [Eq. (\ref{Eq4})] with respect to the electron per undulator period. Taking the two consequences into consideration, we can determine the argument of the wake function for an electron at a certain passage along the undulator on a certain turn.

\begin{figure}
\centering
\includegraphics[width=5in]{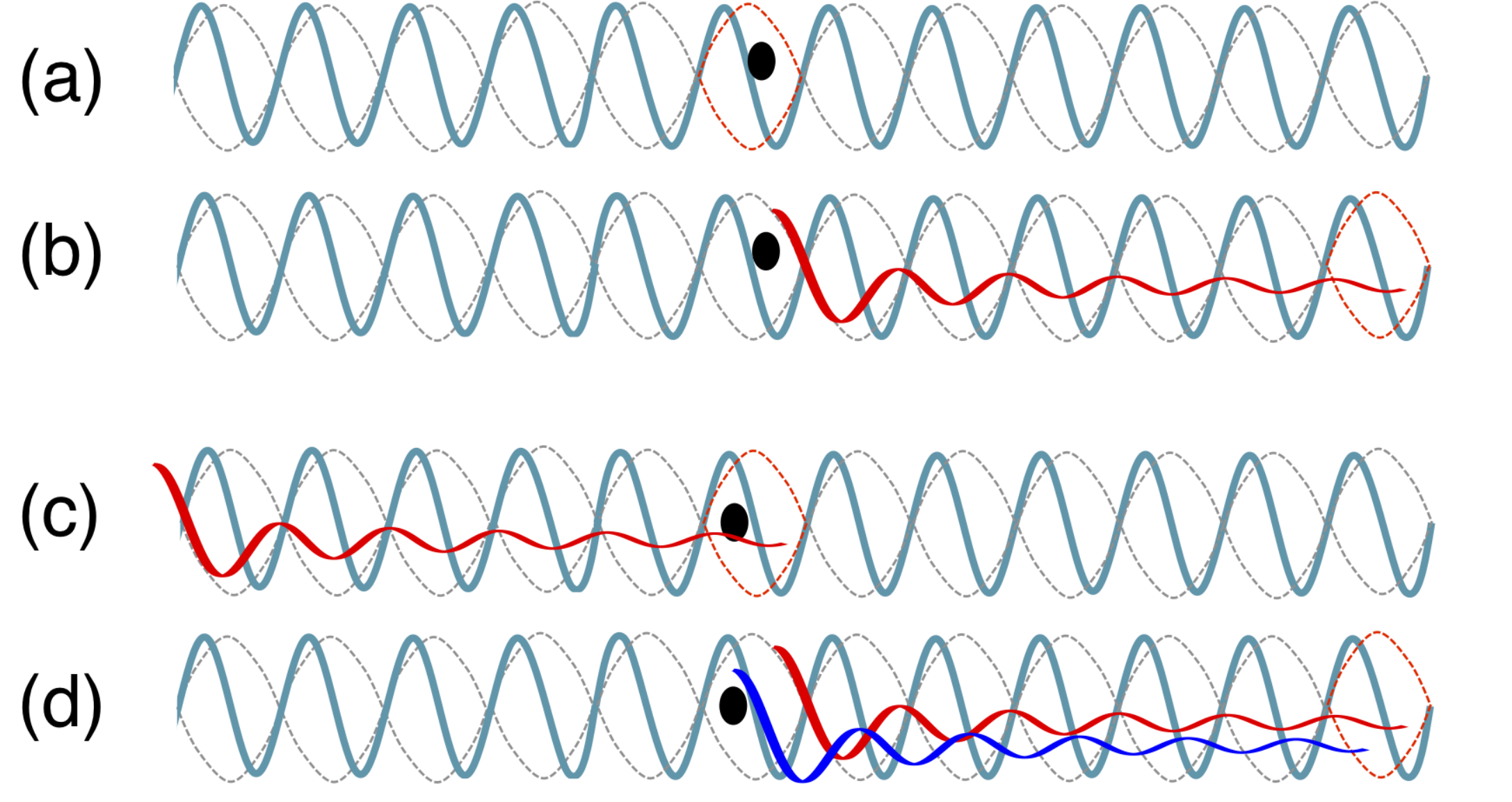}
\caption{\textmd{Illustration of the relative positions of the microbunch to the phase space bucket. (a) Initially on $k = 0$ at the undulator entrance the microbunch lies in the red bucket; (b) Still on $k = 0$ at the undulator exit the microbunch produced radiation wake function and the phase space bucket slips ahead; (c) On $k = 1$ at the undulator entrance after circulating through one turn in the storage ring, the microbunch sees the same phase space bucket (marked in red), together with the radiation wake; (d) On $k = 1$ at the undulator exit the old radiation wake slips ahead, while a new radiation wake (marked in blue) is generated.}}
\label{Fig3}
\end{figure}

To better illustrate the overtaking picture between the circulating electron bunch and the radiation wake, Fig.~\ref{Fig3} shows some specific situations. Let us consider an electron bunch stays in a phase space bucket (marked as red) at the undulator entrance on the very first turn, shown in Fig.~\ref{Fig3}(a). When the bunch traverses one undulator period, it emits radiation. Upon the electron’s leaving from the undulator, the radiation wake takes $N_w\lambda_r$ long because of the slippage effect. Figure~\ref{Fig3}(b) shows a case of $N_w = 6$ (see also Fig.~\ref{Fig1}). After the electron leaves the undulator, it does no longer emit radiation and the generated radiation fields are confined in the laser modulator cavity, for example, the four-mirror ring cavity shown in Fig.~\ref{Fig1}. The phase-locked condition ensures that the electron bunch would meet the same red-marked phase space bucket when returning to the undulator entrance; see Fig.~\ref{Fig3}(c). Possible synchrotron oscillation of the bunch in the remaining storage ring is illustrated. Note that we have assumed both the radiation field and the external laser travel at the same speed $c$ in the modulator cavity. Therefore the electron bunch should see the radiation wake as shown in Fig.~\ref{Fig3}(c): the head of the radiation field is attached to the red-marked phase space bucket. Finally, when the electron bunch travels along the undulator, it consecutively samples the radiation field from head to tail and in the meanwhile produces a new radiation field (marked as blue). Based on the above, we can now write down the functional form of the radiation wakes the circulating electron bunch sees at the $n$-th undulator period on the $m$-th turn
\begin{equation}\label{Eq11}
\sum\limits_{k = 0}^{m - 1} {\sum\limits_{p = 0}^{n - 1} {{{\mathcal{W}}_\parallel }\left[ {\left( {{N_w} - p} \right){\lambda _r} + z_m^{(p)} - z_k^{(0)}} \right]} }, 
\end{equation}
where $\mathcal{W}_{\parallel}$ is denoted as the scaled wake function. The explicit expression will be given in the next subsection [Eq. (\ref{Eq14})]. Here we note that the concept of the phase lock is reflected in the $z_{k}^{(0)}$ of Eq. (\ref{Eq11}). We also note that the radiation wake functions are with a finite duration, i.e., ${{\mathcal{W}}_\parallel }\left( {z < 0} \right) = 0$ and ${{\mathcal{W}}_\parallel }\left( {z > {N_w}{\lambda _r}} \right) = 0$. 

Before ending this subsection we comment on possible effects of cavity mirrors on the radiation fields. From Eq. (\ref{Eq10}) for the case with $\sigma_{\text{refl}} \to \infty$, the field amplitudes decay with a multiplicative factor $R < 1$. The undulator wake functions in the cavity modulator for multiple turns may be modified in the following way
\begin{equation}\label{Eq12}
\sum\limits_{k = 0}^{m - 1} {{R^{m - k}}\sum\limits_{p = 1}^{{N_w} - 1} {{{\mathcal{W}}_\parallel }\left[ {\left( {{N_w} - p} \right){\lambda _r} + z_m^{(p)} - z_k^{(0)}} \right]} }, 
\end{equation}
with the multiplicative factor $\sqrt{R} = \sqrt{1-\alpha}$ with $\alpha$ the power loss. If we include the frequency filtering effect, we may multiply Eq. (\ref{Eq3}) with a Gaussian weighting function ${e^{ - \frac{{{\omega ^2}}}{{4\sigma _{{\text{refl}}}^2}}}}$, and perform an inverse Fourier transformation based on Eq. (\ref{Eq1}) to obtain the wake function. {Here we remark that the absence of explicit $k$ dependence in the argument of the wake function is reflected partially in the fact that on each passage through the modulator undulator the pulse duration of the emitted radiation field is about $N_w\lambda_r$, only occupied in a very limited region of the round-trip distance of the optical cavity. Including the term involving the optical path length in the wake function argument apparently makes the electron beam slip far away from the radiation pulse, thus leading to ineffective laser modulation. In formulating the wake function, the cavity loss effects are included in a multiplicative factor $R$, with $\left|R\right|\le1$. Neglecting the decay effects indeed will result in overestimated interaction between the radiation and electron bunch when they meet on each turn.} Despite all that, in what follows we only focus on the instability mechanism itself by assuming $R = 1, \sigma_{\text{refl}} \to \infty$ and neglecting any effect from the cavity mirrors.

\subsection{Macroparticle equations of motion}\label{SecIIC}
Having obtained the functional form of the radiation wake function Eq. (\ref{Eq11}), we may construct the equations of motion of a circulating microbunch. Although using the macroparticle model to represent a microbunch can only reflect the motion of the bunch centroid, it shall capture the physical essence and serves as the first step for the analysis. Inside the undulator, let us denote as the period-by-period mapping of the longitudinal phase space coordinate from the undulator entrance {$\left( z_m^{(0)}, \delta_m^{(0)} \right)$} to the exit {$\left( z_m^{(N_w)}, \delta_m^{(N_w)} \right)$}, with the subscripts $m$ being the revolution index and the superscript indicating the location of the undulator in terms of undulator period. The superscript $(0)$ refers to the undulator entrance and $(N_w)$ corresponds to the undulator exit. On the $m$-th turn, the equation of motion for the longitudinal displacement at the $(n+1)$-th undulator period can be expressed as~\cite{Ref08}
\begin{equation}\label{Eq13}
z_m^{(n + 1)} = z_m^{(n)} - {\eta _w}{\lambda _w}\delta _m^{(n + 1)}
\end{equation}
with the undulator slippage factor ${\eta _w} =  - {{\left( {1 + \tfrac{{{K^2}}}{2}} \right)} \mathord{\left/
 {\vphantom {{\left( {1 + \tfrac{{{K^2}}}{2}} \right)} {\gamma _0^2}}} \right.
 \kern-\nulldelimiterspace} {\gamma _0^2}}$ and that of the relative energy deviation $\delta = (\gamma-\gamma_0)/\gamma_0$
\begin{equation}\label{Eq14}
\delta _m^{(n + 1)} = \delta _m^{(n)} + \frac{{k_{s0}^2{\lambda _w}}}{{{\eta _w}{k_L}}}\sin {k_L}z_m^{(n)} - \sum\limits_{k = 0}^{m - 1} {{{\mathcal{W}}_\parallel }\left[ {\left( {{N_w} - n} \right){\lambda _r} + z_m^{(n)} - z_k^{(0)}} \right]} 
\end{equation}
with $k_L = 2\pi/\lambda_L$ the laser wavenumber. The scaled wake function is defined as ${{\mathcal{W}}_\parallel } \equiv \tfrac{{4\pi {\epsilon _0}N{r_e}{\lambda _w}}}{{{\gamma _0}}}{W_\parallel }$, where $r_e$ is the classical electron radius. Further, we have ${{\mathcal{W}}_\parallel }\left( {z < 0} \right) = {{\mathcal{W}}_\parallel }\left( {z > {N_w}{\lambda _r}} \right) = 0$. For the case of SSMB, the bucket width $2\pi/k_L \gg \sigma_z$, thus $\sin k_L z_m^{(n)} \approx k_L z_m^{(n)}$.

When the electron bunch leaves the undulator, let us define $\left( {{z_{m,{\text{fin}}}},{\delta _{m,{\text{fin}}}}} \right) = \left( {z_m^{\left( {{N_w}} \right)},\delta _m^{\left( {{N_w}} \right)}} \right)$. The phase space mapping from the undulator exit to the entrance of the next turn $\left( {{z_{m,{\text{fin}}}},{\delta _{m,{\text{fin}}}}} \right) \to \left( {{z_{m + 1,{\text{ini}}}},{\delta _{m + 1,{\text{ini}}}}} \right) = \left( {z_{m + 1}^{\left( 0 \right)},\delta _{m + 1}^{\left( 0 \right)}} \right)$ can be formulated using the linear matrix
\begin{equation}\label{Eq15}
\left[ \begin{gathered}
  {z_{m + 1,{\text{ini}}}} \hfill \\
  {\delta _{m + 1,{\text{ini}}}} \hfill \\ 
\end{gathered}  \right] = \left( {\begin{array}{*{20}{c}}
  1&{ - {\eta _{{\text{ring}}}}\left( {{C_{{\text{tot}}}} - {L_w}} \right)} \\ 
  0&1 
\end{array}} \right)\left[ \begin{gathered}
  {z_{m,{\text{fin}}}} \hfill \\
  {\delta _{m,{\text{fin}}}} \hfill \\ 
\end{gathered}  \right],
\end{equation}
where the total circumference $C_{\text{tot}}$ involves the modulator and the remaining storage ring, $C_{\text{tot}} = L_w + C_{\text{ring}}$.

According to an option of the preliminary SSMB designs, a storage ring with longitudinal strong focusing may be implemented as a quasi-isochronous ring, in which the phase slippage factor $\eta_{\text{ring}}$ of the storage ring itself may be two orders of magnitude smaller than that of the conventional storage rings. For a quasi-isochronous ring, more dedicated studies may include higher order effects of the momentum compaction factor. We also neglect the effects from the transverse chromaticity, i.e., assuming $\Delta z = \alpha C\delta  \gg  - 2\pi \left( {{J_x}{\xi _x} + {J_y}{\xi _y}} \right)$, with $J_{x,y}$ being the single-particle emittance and $\xi{x,y}$ the transverse chromaticities~\cite{Ref33}. In particle tracking simulations the higher order slippage factors can be included by modifying the equation of the longitudinal displacement [Eq. (\ref{Eq15})] 
\begin{equation}\label{Eq16}
{z_{m + 1,{\text{ini}}}} = {z_{m,{\text{fin}}}} - \left( {{C_{{\text{tot}}}} - {L_w}} \right)\left( {{\eta _{{\text{ring}},0}}{\delta _{m,{\text{fin}}}} + {\eta _{{\text{ring}},1}}\delta _{m,{\text{fin}}}^2 +  \cdots } \right).
\end{equation}

In the following discussion however we only focus on the instability mechanism and only retain the first order term $\eta_{\text{ring,0}}$. We have thus far obtained two sets of longitudinal equations of motion, one for the undulator Eqs. (\ref{Eq13},\ref{Eq14}) and the other for the remaining storage ring Eq. (\ref{Eq15}). Below we will simply these equations by making two assumptions. First, because the first term $(N_w - p)\lambda_r$ in the argument of the wake functions in Eq. (\ref{Eq11}) is typically much larger than the second term $(z_m^{(p)}) – z_k^{(0)}$, we may Taylor expand
\begin{equation}\label{Eq17}
{{\mathcal{W}}_\parallel }\left[ {\left( {{N_w} - p} \right){\lambda _r} + z_m^{(p)} - z_k^{(0)}} \right] \approx {{\mathcal{W}}_\parallel }\left[ {\left( {{N_w} - p} \right){\lambda _r}} \right] + \left( {z_m^{(p)} - z_k^{(0)}} \right){{\mathcal{W}}_\parallel^{\prime} }\left[ {\left( {{N_w} - p} \right){\lambda _r}} \right],
\end{equation}
where the first term on the right hand side is the parasitic loss, a static term independent of the beam motion, and will be dropped in the analysis. The second term proportional to $z_m^{(p)}$ refers to the potential well distortion effect~\cite{Ref09,Ref10}. Of our most interest is the third term proportional to $z_m^{(p)}$, which characterizes the driving source and participates the collective dynamics. Second, we define the undulator-averaged phase space coordinates as
\begin{equation}\label{Eq18}
{\bar z_m} = \frac{1}{{{N_w}}}\sum\limits_{n = 1}^{{N_w}} {z_m^{(n)}} ,{\bar \delta _m} = \frac{1}{{{N_w}}}\sum\limits_{n = 1}^{{N_w}} {\delta _m^{(n)}}. 
\end{equation}
Here it does not matter to take average of the undulator periods from 0 to $N_w -1$ or from $1$ to $N_w$. Now we want to simplify Eqs. (\ref{Eq13})-(\ref{Eq15}) using Eq. (\ref{Eq17}) and (\ref{Eq18}). The radiation wake sum in Eq. (\ref{Eq14}) has the simplified form
\begin{equation}\label{Eq19}
\sum\limits_{p = 0}^{{N_w} - 1} {\left( {z_m^{(p)} - z_k^{(0)}} \right){{{\mathcal{W}}}_\parallel^{\prime} }\left[ {\left( {{N_w} - p} \right){\lambda _r}} \right]}  \approx \left( {\bar z_m^{} - \bar z_k^{}} \right)\sum\limits_{p = 0}^{{N_w} - 1} {{{{\mathcal{W}}}_\parallel^{\prime} }\left[ {\left( {{N_w} - p} \right){\lambda _r}} \right]}  \equiv \left( {\bar z_m^{} - \bar z_k^{}} \right){\mathbb{W}'},
\end{equation}
where in the last equality a shorthand notation $\mathbb{W}^{\prime}$ is introduced. The explicit expressions of $\mathbb{W}^{\prime}$ with detailed derivations can be found in Appendix A. It is worth noting that $\mathbb{W}^{\prime}$ in Eq. (\ref{Eq19}) does not have $m$-dependence after the sum over $p$. This is expected because the information of the radiation fields generated by a turn-by-turn circulating electron was contained in $\bar{z}_k$ [see also Eq. (\ref{Eq18})]. {In the approximate expression we have assumed that the longitudinal bunch centroid does not change too much during its passage through the modulator undulator. In this situation the quantity $z$ can be moved out from the summation over $p$ and denoted as $\bar{z}$. This approximation is valid only when the synchrotron motion is not evident inside the modulator undulator.} This functional form of the radiation wake sum in Eq. (\ref{Eq19}) is different from that of the Robinson instability~\cite{Ref09}. For such a difference we will briefly discuss in Sec.~\ref{SecV}. Using Eqs. (\ref{Eq17}-\ref{Eq19}), the two sets of equations of motion become
\begin{equation}\label{Eq20}
\frac{{{\text{d}}{{\bar z}_m}}}{{{\text{d}}m}} =  - \left( {{\eta _w}{L_w} + {\eta _{{\text{ring}}}}{C_{{\text{ring}}}}} \right){\bar \delta _m}
\end{equation}
with ${C_{{\text{ring}}}} = {C_{{\text{tot}}}} - {L_w}$ and
\begin{equation}\label{Eq21}
\frac{{{\text{d}}{{\bar \delta }_m}}}{{{\text{d}}m}} = \frac{{k_{s0}^2{L_w}}}{{{\eta _w}}}{\bar z_m} - {\mathbb{W}'}\sum\limits_{k = 0}^{m - 1} {\left( {{{\bar z}_m} - {{\bar z}_k}} \right)} 
\end{equation}
or, combined as a second-order difference equation
\begin{equation}\label{Eq22}
\frac{{{{\text{d}}^2}{{\bar z}_m}}}{{{\text{d}}{m^2}}} + 4{\pi ^2}\nu _{s0,{\text{tot}}}^2{\bar z_m} = \left( {{\eta _w}{L_w} + {\eta _{{\text{ring}}}}{C_{{\text{ring}}}}} \right){\mathbb{W}'}\sum\limits_{k = 0}^{m - 1} {\left( {{{\bar z}_m} - {{\bar z}_k}} \right)}, 
\end{equation}
in which the longitudinal synchrotron oscillation tune for the entire ring
\begin{equation}\label{Eq23}
\nu _{s0,{\text{tot}}}^2 = \frac{{k_{s0}^2L_w^2}}{{4{\pi ^2}}}\frac{{{\eta _w}{L_w} + {\eta _{{\text{ring}}}}{C_{{\text{ring}}}}}}{{{\eta _w}{L_w}}}.
\end{equation}
Here we note that $k_{s0}$ is provided by the external laser and can be written as~\cite{Ref08}
\begin{equation}\label{Eq24}
{k_{s0}} = \sqrt {\frac{{e{{\mathcal{E}}_0}K[JJ]{k_w}}}{{\gamma _L^2m{c^2}}}},
\end{equation}
with $\left[ {JJ} \right] = {J_0}\left( \chi  \right) - {J_1}\left( \chi  \right),\chi  = \tfrac{{{K^2}}}{{4 + 2{K^2}}}$, the electric field amplitude ${{\mathcal{E}}_0} \approx {{2\gamma {V_m}} \mathord{\left/ {\vphantom {{2\gamma {V_m}} {K{L_w}}}} \right. \kern-\nulldelimiterspace} {K{L_w}}}$. The $\gamma_L$ can be related to the laser wavelength and the undulator parameters satisfying a similar relation to Eq. (\ref{Eq4}). One can also see from Eq. (\ref{Eq22}) that in the absence of radiation field, the single-particle stability requires ${\eta _w}{L_w} + {\eta _{{\text{ring}}}}{C_{{\text{ring}}}} < 0$. For an undulator, the slippage factor $\eta_w = -\frac{1+K^2/2}{\gamma_0^2} < 0$.

The above Eqs. (\ref{Eq20}) and (\ref{Eq21}) can also be written in a matrix form
\begin{equation}\label{Eq25}
\left[ \begin{gathered}
  {{\bar z}_{m + 1}} \hfill \\
  {{\bar \delta }_{m + 1}} \hfill \\ 
\end{gathered}  \right] \approx \left( {\begin{array}{*{20}{c}}
  1&{ - {\eta _{{\text{ring}}}}{C_{{\text{ring}}}}} \\ 
  0&1 
\end{array}} \right)\left\{ {\left( {\begin{array}{*{20}{c}}
  1&{ - {\eta _w}{L_w}} \\ 
  {\tfrac{{k_{s0}^2{L_w}}}{{{\eta _w}}} - m{\mathbb{W}'}}&1 
\end{array}} \right)\left[ \begin{gathered}
  {{\bar z}_m} \hfill \\
  {{\bar \delta }_m} \hfill \\ 
\end{gathered}  \right] + \left[ \begin{gathered}
  0 \hfill \\
  {\mathbb{W}'}\sum\limits_{k = 0}^{m - 1} {{{\bar z}_k}}  \hfill \\ 
\end{gathered}  \right]} \right\},
\end{equation}
where the term $m{\mathbb{W}'}$ comes from the potential well distortion effect.

We have at the moment derived the linearized equations of motion in the laser modulator cavity [Eqs. (\ref{Eq13},\ref{Eq14})] and in the remaining storage ring [Eq. (\ref{Eq15})]. We have also obtained the simplified equations of motion Eqs. (\ref{Eq20}-\ref{Eq22}), after defining the undulator-averaged phase space coordinates. In Sec.~\ref{SecIII} we will proceed with solving the equations for system instability.

\subsection{Pure optics single particle motion}\label{SecIID}
Before proceeding with the instability analysis, let us pause for a moment and consider the single-particle optics in the absence of radiation fields. The evolution of an electron traversing the undulator and circulating in the remaining storage ring can be formulated using the respective transport matrices $M_{L_w}$ and $M_{C_{\text{ring}}}$. The one-turn matrix can be written as~\cite{Ref36,Ref37}
\begin{align}\label{Eq26}
OTM &= {M_{{C_{{\text{ring}}}}}}{M_{{L_w}}} = \left( {\begin{array}{*{20}{c}}
  1&{ - {\eta _{{\text{ring}}}}\left( {{C_{{\text{tot}}}} - {L_w}} \right)} \\ 
  0&1 
\end{array}} \right)\left( {\begin{array}{*{20}{c}}
  {\cos {k_{s0}}{L_w}}&{ - \frac{{{\eta _w}}}{{{k_{s0}}}}\sin {k_{s0}}{L_w}} \\ 
  {\frac{{{k_{s0}}}}{{{\eta _w}}}\sin {k_{s0}}{L_w}}&{\cos {k_{s0}}{L_w}} 
\end{array}} \right) \nonumber \\
&= \left( {\begin{array}{*{20}{c}}
  {\cos {k_{s0}}{L_w} - \frac{{{\eta _{{\text{ring}}}}}}{{{\eta _w}}}\left( {\frac{{{C_{{\text{tot}}}}}}{{{L_w}}} - 1} \right){k_{s0}}{L_w}\sin {k_{s0}}{L_w}}&{ - \frac{{{\eta _w}}}{{{k_{s0}}}}\sin {k_{s0}}{L_w} - {\eta _{{\text{ring}}}}\left( {{C_{{\text{tot}}}} - {L_w}} \right)\cos {k_{s0}}{L_w}} \\ 
  {\frac{{{k_{s0}}}}{{{\eta _w}}}\sin {k_{s0}}{L_w}}&{\cos {k_{s0}}{L_w}} 
\end{array}} \right),
\end{align}
where this mapping is symplectic. It can be seen that when ${k_{s0}}{L_w} \ll 1$, the above is reduced to Eq. (\ref{Eq25}) in the absence of collective effects. Therefore Eq. (\ref{Eq25}) is valid when ${k_{s0}}{L_w} \ll 1$.

The single-particle stability can be determined from Eq. (\ref{Eq26}) by requiring $|\text{Tr} OTM| < 2$, giving the exact synchrotron tune of the total storage ring
\begin{align}\label{Eq27}
{\nu _{s0,{\text{tot,EXACT}}}} &= \frac{1}{{2\pi }}{\cos ^{ - 1}}\left( {\frac{{{\text{Tr }}OTM}}{2}} \right) \nonumber \\
&= \frac{1}{{2\pi }}{\cos ^{ - 1}}\left[ {\cos {k_{s0}}{L_w} - \frac{{{\eta _{{\text{ring}}}}}}{{{\eta _w}}}\left( {\frac{{{C_{{\text{tot}}}}}}{{{L_w}}} - 1} \right)\frac{{{k_{s0}}{L_w}}}{2}\sin {k_{s0}}{L_w}} \right],
\end{align}
where ${\nu _{s0}} = \frac{{{k_{s0}}{L_w}}}{{2\pi }} = \frac{{{k_{s0}}}}{{{{2\pi } \mathord{\left/
 {\vphantom {{2\pi } {{L_w}}}} \right.
 \kern-\nulldelimiterspace} {{L_w}}}}}$. For ${k_{s0}}{L_w} \ll 1$, we have
\begin{equation}\label{Eq28}
{\nu _{s0,{\text{tot,EXACT}}}} \approx {\nu _{s0,{\text{tot}}}} = \frac{{{k_{s0}}{L_w}}}{{2\pi }}\sqrt {1 + \frac{{{\eta _{{\text{ring}}}}}}{{{\eta _w}}}\frac{{{C_{{\text{tot}}}} - {L_w}}}{{{L_w}}}}, 
\end{equation}
which is the obtained Eq. (\ref{Eq23}). 

\begin{figure}
\centering
\includegraphics[width=4in]{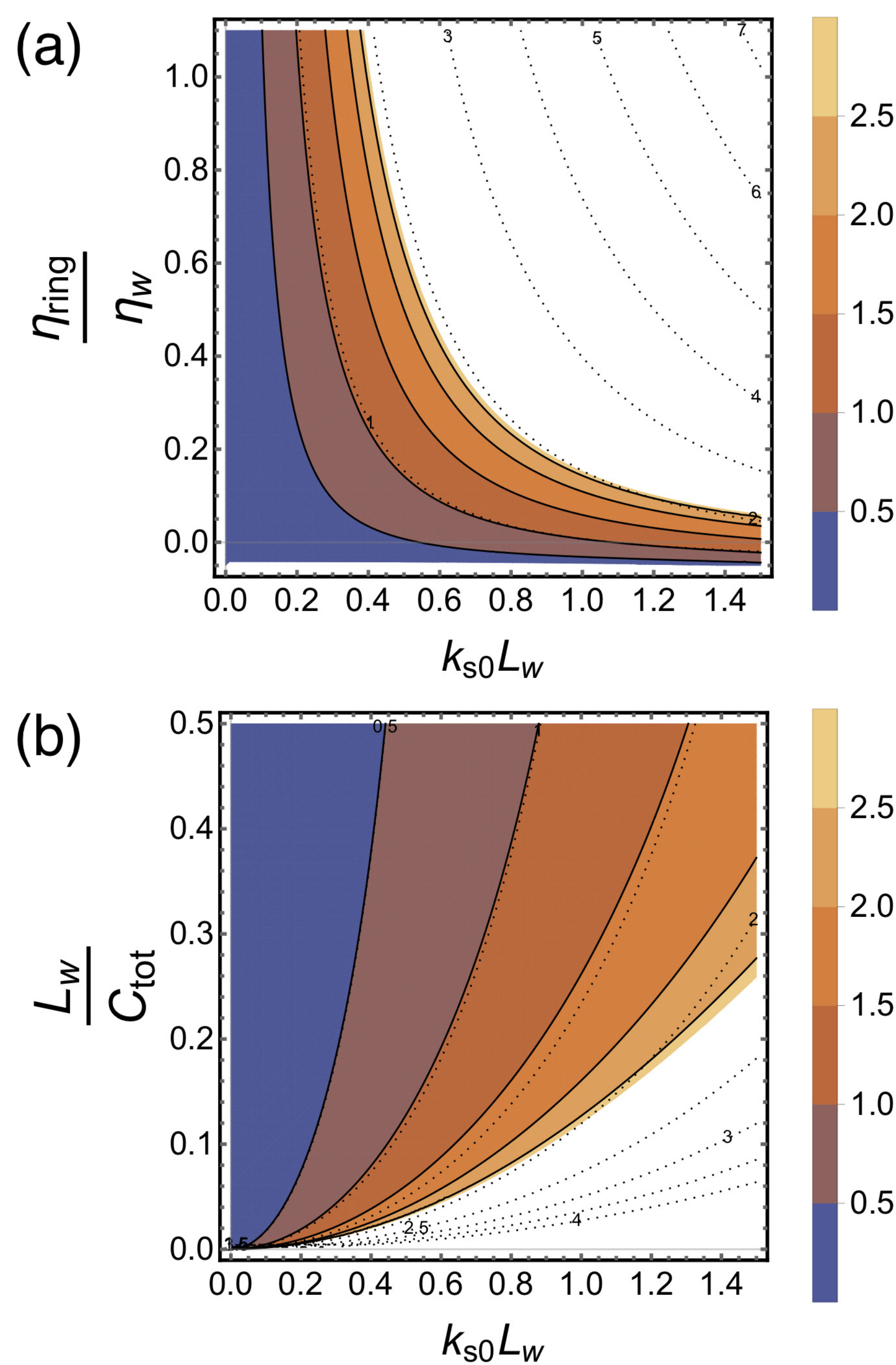}
\caption{\textmd{Single-particle stability diagram for (a) ${L_w} = 2{\text{ m, }}{C_{{\text{tot}}}} - {L_w} = 48{\text{ m}}$; (b) $\eta _{\text{ring}}/\eta _w = 0.5$. The color bar refers to $\nu _{s,{\text{tot,EXACT}}}/\nu _{s0{\text{,tot}}}$. The solid contour lines are based the exact expression [Eq. (27)]; the dashed lines based on the approximate expression [Eq. (28)]. The white, unshaded regions are the unstable motion.}}
\label{Fig4}
\end{figure}

Figure~\ref{Fig4} gives the ratio ${{{\nu _{s,{\text{tot,EXACT}}}}} \mathord{\left/
 {\vphantom {{{\nu _{s,{\text{tot,EXACT}}}}} {{\nu _{s0{\text{,tot}}}}}}} \right.
 \kern-\nulldelimiterspace} {{\nu _{s0{\text{,tot}}}}}}$ [see Eqs. (\ref{Eq27}) and (\ref{Eq28})] as functions of $\eta_{\text{ring}}/\eta_{w}$ and $L_w/C_{\text{tot}}$. One can find that, the stronger focusing, i.e., the larger $k_{s0}L_w$, may lead to the more restricted single-particle stability region. Further, the ratio ${{{\nu _{s,{\text{tot,EXACT}}}}} \mathord{\left/
 {\vphantom {{{\nu _{s,{\text{tot,EXACT}}}}} {{\nu _{s0{\text{,tot}}}}}}} \right.
 \kern-\nulldelimiterspace} {{\nu _{s0{\text{,tot}}}}}}$ also depends on the designs of the modulator and the storage ring.

\begin{figure}
\centering
\includegraphics[width=4in]{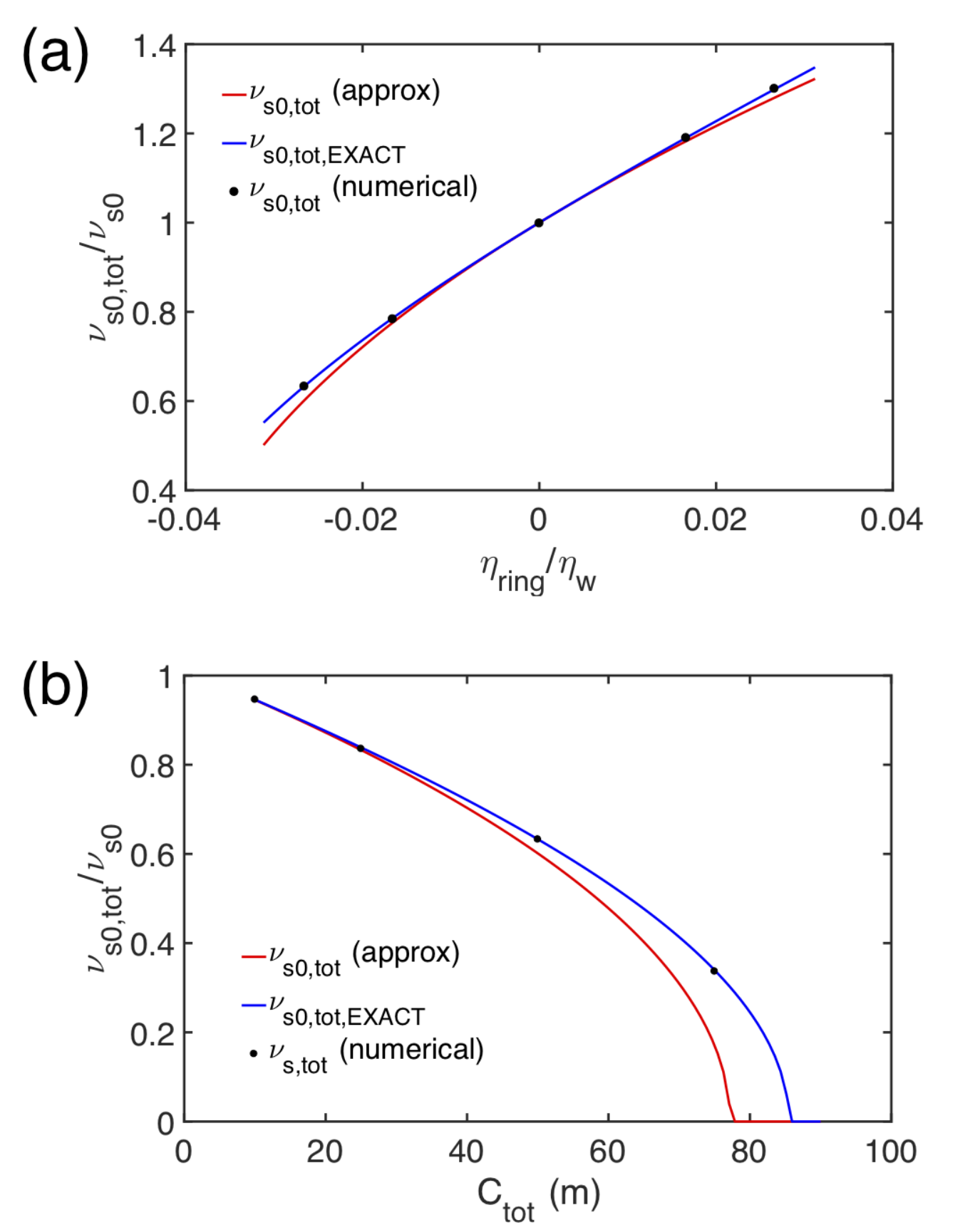}
\caption{\textmd{$\nu_{s0,\text{tot}}/\nu_{s0}$ based on the exact, approximate, and numerical methods as a function of (a) $\eta_{\text{ring}}/\eta_w$ with ${L_w} = 2{\text{ m, }}{C_{{\text{tot}}}} - {L_w} = 48{\text{ m}}$; and (b) the total circumference given $\eta _{{\text{ring}}}/ \eta _w =  - 0.0266$.}}
\label{Fig5}
\end{figure}

Turn-by-turn macroparticle tracking simulation can be implemented by numerically iterating Eqs. (\ref{Eq13}-\ref{Eq15}) or Eqs. (\ref{Eq20},\ref{Eq21}). The synchrotron oscillation frequencies can be extracted by Fourier transformation of the turn-by-turn $\bar{z}_m$, as shown in Fig.~\ref{Fig5}. Figure~\ref{Fig5}(a) shows the dependence of $\eta_{\text{ring}}/\eta_{w}$ on the synchrotron tune for a fixed ring-undulator size ${L_w} = 2{\text{ m, }}{C_{{\text{tot}}}} - {L_w} = 48{\text{ m}}$. Under a specific configuration of ${{{\eta _{{\text{ring}}}}} \mathord{\left/ {\vphantom {{{\eta _{{\text{ring}}}}} {{\eta _w}}}} \right. \kern-\nulldelimiterspace} {{\eta _w}}} =  - 0.0266$, Fig.~\ref{Fig5}(b) gives how the synchrotron tune varies with the total circumference. The turn-by-turn tracking simulations confirm that Eq. (\ref{Eq27}) is a more accurate prediction and only when the phase slippage factor or the total circumference is small, does the approximate expression Eq. (\ref{Eq28}) give a good prediction. In Fig.~\ref{Fig5}(b), when exceeding the cutoff, the particle would not execute the synchrotron oscillation due to the stability criterion.

\section{Dispersion equation and instability growth rate}\label{SecIII}
In Sec.~\ref{SecIIC} we have obtained the equations of motion in the laser modulator cavity [Eqs. (\ref{Eq13},\ref{Eq14})] and in the remaining storage ring [Eq. (\ref{Eq15})]. After introducing the undulator-averaged phase space coordinates, we have derived the simplified equations of motion Eqs. (\ref{Eq20}-\ref{Eq22}) or Eq. (\ref{Eq25}). Now we want to analyze the stability of such a dynamical system by finding the dispersion equation, and if unstable, determine the instability growth rate. To solve for the difference equations we shall apply z-transform~\cite{Ref38}. For convenience of the subsequent analyses, let us define the $m$-th turn phase space vector $Y_m = \left[ \bar{z}_m \text{ } \bar{\delta}_m \right]^T$ with its z-transform
\begin{equation}\label{Eq29}
{\mathcal{Z}}\left[ {{Y_m}} \right] = \tilde Y(\mathsf{z}) \equiv \sum\limits_{j = 0}^\infty  {{Y_j}{\mathsf{z}^{ - j}}}  = {Y_0} + {Y_1}{\mathsf{z}^{ - 1}} + {Y_2}{\mathsf{z}^{ - 2}} +  \cdots, 
\end{equation}
where $\mathsf{z}$ is in general a complex quantity. Such a transformation is similar to the Laplace transform in a continuous or differential equation. Neglecting the potential well distortion effect in Eq. (\ref{Eq25}), the matrix equation becomes
\begin{equation}\label{Eq30}
\left[ \begin{gathered}
  {{\bar z}_{m + 1}} \hfill \\
  {{\bar \delta }_{m + 1}} \hfill \\ 
\end{gathered}  \right] \approx \left( {\begin{array}{*{20}{c}}
  1&{ - \left( {{\eta _w}{L_w} + {\eta _{{\text{ring}}}}{C_{{\text{ring}}}}} \right)} \\ 
  {\tfrac{{k_{s0}^2{L_w}}}{{{\eta _w}}}}&1 
\end{array}} \right)\left[ \begin{gathered}
  {{\bar z}_m} \hfill \\
  {{\bar \delta }_m} \hfill \\ 
\end{gathered}  \right] + \left( {\begin{array}{*{20}{c}}
  0&0 \\ 
  {{\mathbb{W}'}}&0 
\end{array}} \right)\sum\limits_{k = 0}^{m - 1} {\left[ \begin{gathered}
  {{\bar z}_k} \hfill \\
  {{\bar \delta }_k} \hfill \\ 
\end{gathered}  \right]}
\end{equation}
or in a more compact form
\begin{equation}\label{Eq31}
{Y_{m + 1}} = {\mathbf{A}}{Y_m} + {\mathbf{B}}\sum\limits_{k = 0}^{m - 1} {{Y_k}}. 
\end{equation}

Notice that Eq. (\ref{Eq31}) is a matrix equation with memory. The source of the memory originates from the radiation wakes. Now we take z-transform on both sides of Eq. (\ref{Eq31}), giving
\begin{equation}\label{Eq32}
\mathsf{z}\tilde Y(\mathsf{z}) - \mathsf{z}{Y_0} = {\mathbf{A}}\tilde Y(\mathsf{z}) + {\mathbf{B}}\frac{{\tilde Y(\mathsf{z})}}{{\mathsf{z} - 1}},
\end{equation}
in which the following two relations are used~\cite{Ref38}
\begin{equation}\label{Eq33}
{\mathcal{Z}}\left[ {{Y_{m + k}}} \right] = {\mathsf{z}^k}\tilde Y(\mathsf{z}) - \sum\nolimits_{r = 0}^{k - 1} {{Y_r}{\mathsf{z}^{k - r}}} 
\end{equation}
and
\begin{equation}\label{Eq34}
{\mathcal{Z}}\left[ {\sum\limits_{k = 0}^{m - 1} {{Y_k}} } \right] = \frac{{\tilde Y(\mathsf{z})}}{{\mathsf{z} - 1}}.
\end{equation}

We solve Eq. (\ref{Eq32}) for $\tilde{Y}$ to obtain
\begin{equation}\label{Eq35}
\tilde Y(\mathsf{z}) = \frac{\mathsf{z}}{{\mathsf{z}\mathbb{I} - {\mathbf{A}} - \frac{{\mathbf{B}}}{{\mathsf{z} - 1}}}}{Y_0},
\end{equation}
with $\mathbb{I}$ the identity matrix.

Note here that the matrix $\mathbf{B}$ is independent of $m$ and does not participate the transformation. The long-term (in)stability is determined by the properties of the denominator of Eq. (\ref{Eq35}), i.e., the dispersion or secular equation $\det \left( {\mathsf{z}\mathbb{I} - {\mathbf{A}} - \frac{{\mathbf{B}}}{{\mathsf{z} - 1}}} \right) = 0$. The dispersion equation can be explicitly expressed as
\begin{equation}\label{Eq36}
\det \left( {\begin{array}{*{20}{c}}
  {\mathsf{z} - 1}&{{\eta _w}{L_w} + {\eta _{{\text{ring}}}}{C_{{\text{ring}}}}} \\ 
  { - \tfrac{{k_{s0}^2{L_w}}}{{{\eta _w}}} - \tfrac{{{\mathbb{W}'}}}{{\mathsf{z} - 1}}}&{\mathsf{z} - 1} 
\end{array}} \right) = 0
\end{equation}
or
\begin{equation}\label{Eq37}
{\left( {\mathsf{z} - 1} \right)^2} + \left( {{\eta _w}{L_w} + {\eta _{{\text{ring}}}}{C_{{\text{ring}}}}} \right)\left( {\tfrac{{k_{s0}^2{L_w}}}{{{\eta _w}}} + \tfrac{{{\mathbb{W}'}}}{{\mathsf{z} - 1}}} \right) = 0.
\end{equation}

Now we let $\mathsf{z} = {e^{i\Omega {T_0}}}$ with $\Omega$ in general a complex quantity. Considering $|\Omega T_0| \ll 1$, we have
\begin{equation}\label{Eq38}
{\Omega ^3} - \omega _{s0,{\text{tot}}}^2\Omega  + i{\hat{\mathbb{W}}} = 0,
\end{equation}
with the synchrotron oscillation frequency [see also Eq. (\ref{Eq23})]
\begin{equation}\label{Eq39}
\omega _{s0,{\text{tot}}}^2 = \frac{{4{\pi ^2}\nu _{s0,{\text{tot}}}^2}}{{T_0^2}}
\end{equation}
and
\begin{equation}\label{Eq40}
{\hat{\mathbb{W}}} = \frac{{{\mathbb{W}'}}}{{T_0^3}}\left( {{\eta _w}{L_w} + {\eta _{{\text{ring}}}}{C_{{\text{ring}}}}} \right).
\end{equation}

Here the explicit expression of $\mathbb{W}^{\prime}$ can be found in Eq. (\ref{EqA5}). Note that $\hat{\mathbb{W}}$ is a real quantity. To solve Eq. (\ref{Eq38}), we define another shorthand notation
\begin{equation}\label{Eq41}
\Delta  = \sqrt {81{{{\hat{\mathbb{W}}}}^2} + 12\omega _{s0,{\text{tot}}}^6}  - 9{\hat{\mathbb{W}}}.
\end{equation}

The three solutions to Eq. (\ref{Eq38}) can be expressed as
\begin{align}\label{Eq42}
{\Omega _0} &= \left[ {{{\left( {\frac{{\sqrt 3 }}{{4\Delta }}} \right)}^{\frac{1}{3}}}\omega _{s0,{\text{tot}}}^2 + {{\left( {\frac{\Delta }{{16\sqrt 3 }}} \right)}^{\frac{1}{3}}}} \right] + \frac{i}{{\sqrt 3 }}\left[ { - {{\left( {\frac{{\sqrt 3 }}{{4\Delta }}} \right)}^{\frac{1}{3}}}\omega _{s0,{\text{tot}}}^2 + {{\left( {\frac{\Delta }{{16\sqrt 3 }}} \right)}^{\frac{1}{3}}}} \right] \nonumber \\
&= {\Omega _{0,R}} + i{\Omega _{0,I}},
\end{align}

\begin{equation}\label{Eq43}
{\Omega _1} = {\Omega _{1,R}} + i{\Omega _{1,I}} =  - {\Omega _{0,R}} + i{\Omega _{0,I}},
\end{equation}
and
\begin{equation}\label{Eq44}
{\Omega _2} = {\Omega _{2,R}} + i{\Omega _{2,I}} = 0 - 2i{\Omega _{0,I}},
\end{equation}
with $\sum\limits_{i = 0}^2 {{\Omega _i}}  = 0$. When $\Delta > 0, \omega_{s0,\text{tot}} > 0$, the real and imaginary parts of the three roots can be formulated in Eqs. (\ref{Eq42}-\ref{Eq44}) using $\Omega_{j,R}$ and $\Omega_{j,I}$. When $\Delta \to 0, \omega_{s0,\text{tot}} \to 0$, all three roots become the trivial solution $\Omega \to 0$. In Eqs. (\ref{Eq42}-\ref{Eq44}) the real part of the solutions characterize the oscillations of the bunch centroid, while the imaginary part refers to the oscillation amplitude growth (or damping). A negative value of the imaginary part indicates the instability growth. Note that the general expression of the solutions does not require $\Delta \ge 0$. When the undulator and the storage ring are designed to satisfy the single-particle stability, we have $\omega_{s0,\text{tot}} > 0$, $\Delta > 0$. We also find that in the absence of any external damping mechanism such a dynamical system is always unstable{, except for $\eta_w L_w + \eta_{\text{ring}}C_{\text{ring}} = 0$}. Those with ${\Omega _{i,I}} < 0$ will lead to instability with a growth rate ${\tau ^{ - 1}} \approx  - {\Omega _{i,I}}$.

\begin{figure}
\centering
\includegraphics[width=4in]{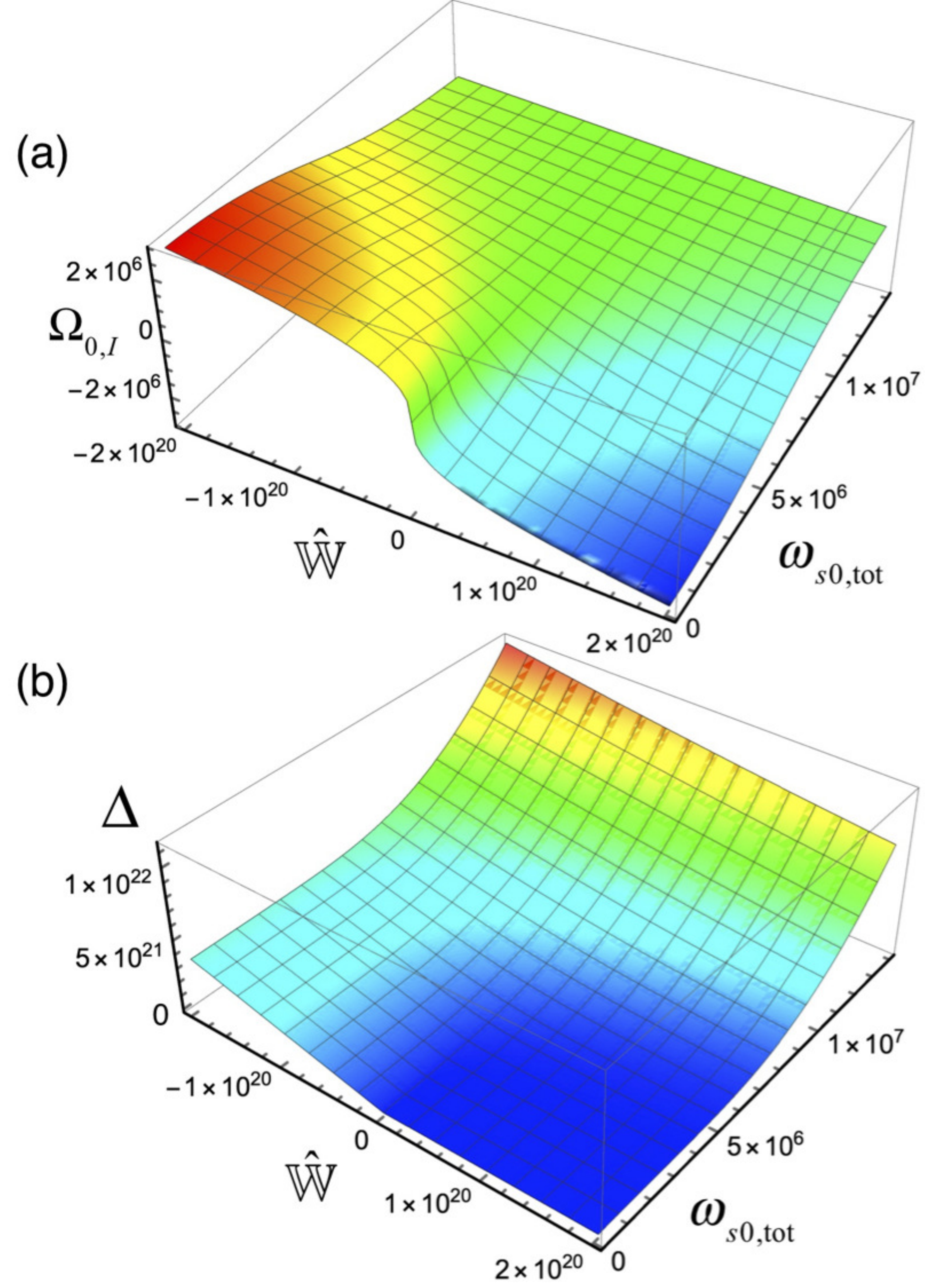}
\caption{\textmd{Parametric dependencies of (a) $\Omega_{0,I}$ (negative values refer to the unstable motion) and (b) $\Delta$ on the synchrotron oscillation frequency $\omega_{s0,\text{tot}}$ and the scaled radiation wake strength $\hat{\mathbb{W}}$.}}
\label{Fig6}
\end{figure}

To illustrate the parametric dependence of the above two quantities $\Omega_{0,I}$ and $\Delta$, Fig. 6 shows their dependence on the scaled radiation wake strength $\hat{\mathbb{W}}$ [Eq. (\ref{Eq39})] and the synchrotron oscillation frequency $\omega_{s0,\text{tot}}$ [Eq. (\ref{Eq40})]. The plots in Fig.~\ref{Fig6} are largely based on the undulator parameters given in Table I (Sec. IV).

In Fig.~\ref{Fig6}(a), the $\Omega_{0,I} < 0$ means an unstable motion. It can be seen that a larger $\omega_{s0,\text{tot}}$, indicating stronger longitudinal focusing, may help mitigate the instability growth due to $\Omega_{0,I}$. Note here that it does not eliminate the instability and fully stabilizes the system, because either of the remaining two solutions has always the opposite sign of $\Omega_{0,I}$. When $\omega_{s0,\text{tot}}$ becomes smaller, e.g., $5 \times 10^{6}$ rad/sec, the sign of $\Omega_{0,I}$ depends on the sign of $\hat{\mathbb{W}}$ [see also Eq. (\ref{Eq40})], i.e., the storage ring design. Here $\omega_{s0,\text{tot}} \approx 5 \times 10^{6}$ rad/sec roughly corresponds to a weaker laser modulator voltage of 0.51 kV. 

The above observations indicate that a strong longitudinal focusing may help mitigate the instability. When the external focusing is not sufficient, smaller instability growth rate may benefit from a proper design of the storage ring and the laser modulator. We will elaborate this in the next section by taking a preliminary SSMB storage ring design parameters as an example.

\section{Example}\label{SecIV}
In this section we illustrate a preliminary example as a practical application of the developed theoretical formulation. This example is designed to a produce high repetition rate, kilowatt-level average power, coherent extreme ultraviolet (EUV) radiation source with the targeted radiation wavelength at 13.5 nm. The preliminary beam and undulator parameters for the example are summarized in Table I. Some of the numbers quoted in Table I have been slightly different from those in Ref.~\cite{Ref04} for illustration purpose. In particular, the bunch charge 6.4 fC adopted here is about an order of magnitude larger than that in Ref.~\cite{Ref04}. Moreover in Table I we presume $\eta_{\text{ring}} = 0$ for the moment. As mentioned in Sec.~\ref{SecIIC}, when $\eta_{\text{ring}}$ is small, the higher order terms should be included. Here and in the following we retain only the linear term and put more emphasis on the instability mechanism.

\begin{table}[]
\caption{Relevant beam and undulator parameters for the SSMB EUV example.}
\begin{tabular}{lcc}
\hline\hline
Name                                                        & Value        & Unit  \\ \hline
Electron energy $\gamma_0 mc^2$         & $400$          & MeV   \\
Bunch charge $Ne$                                  & $6.4$          & fC    \\
Modulator voltage $V_m$                         & $0.51$         & MV    \\
Undulator parameter $K$                         & $4.2$          &       \\
Undulator slippage factor $\eta_w$        	        & $-1.6 \times 10^{-5}$    &       \\
Unperturbed synchrotron wavenumber $k_{s0}$    & $0.54 = 2\pi/11.6$         & rad/m \\
Undulator period $\lambda_w$                                     & $2$            & cm    \\
Resonant radiation wavelength $\lambda_r$              & $0.16$         & $\mu$m    \\
Undulator length $L_w$                                     & $2$            & m     \\
Maximum harmonic number $H_{\text{max}}$                   & $1$            &       \\
Storage ring total circumference $C_{\text{tot}}$             & $50$           & m     \\
Storage ring slippage factor $\eta_{\text{ring}}$                    & $0$ (may vary) &       \\ \hline\hline
\end{tabular}
\end{table}

\begin{figure}
\centering
\includegraphics[width=5in]{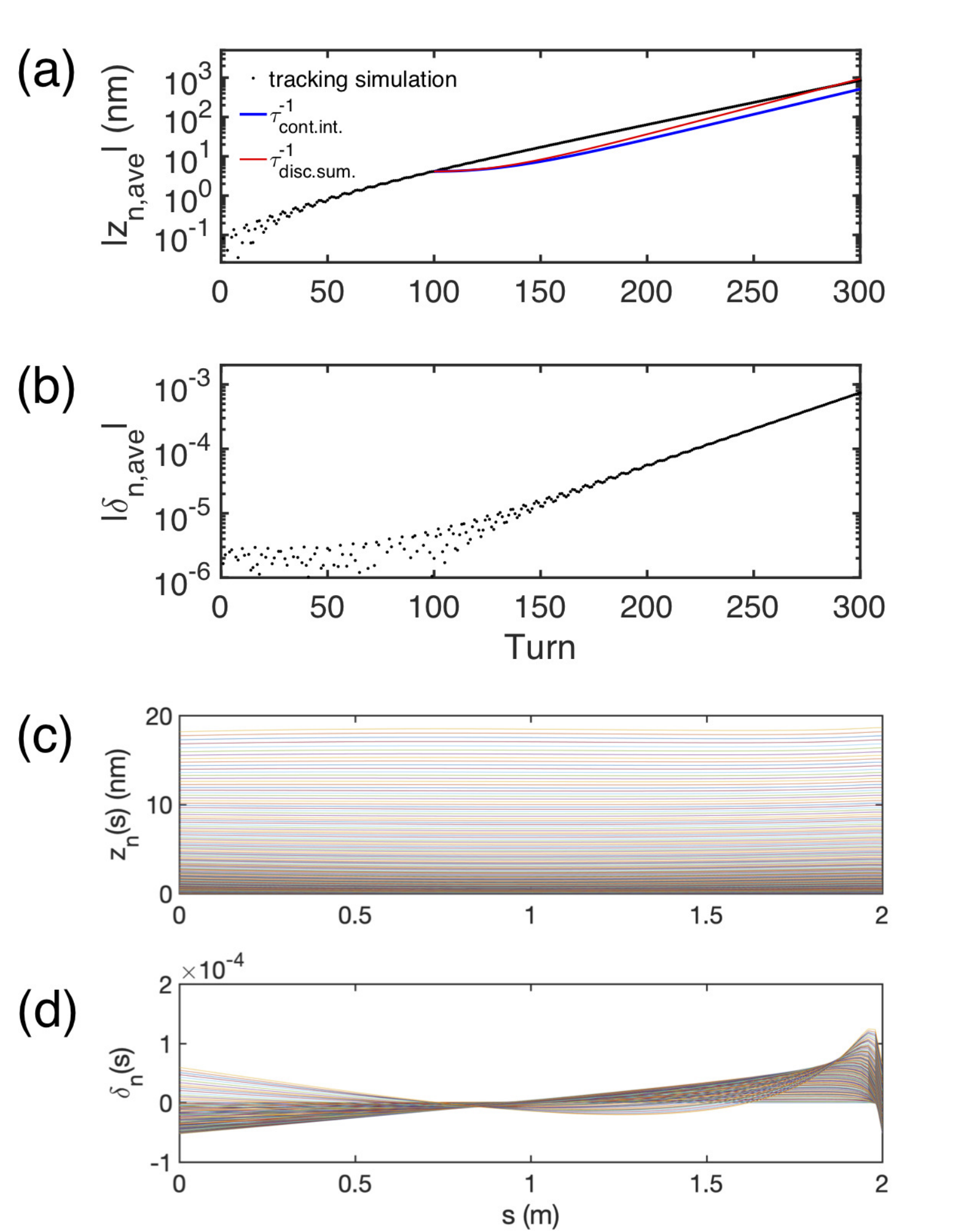}
\caption{\textmd{Evolution of the undulator-averaged phase space coordinates turn-by-turn in the storage ring (a,b) and in the undulator of the laser modulator (c,d). The beam and undulator parameters are given in Table I. Here we use the linearized radiation wake function and exclude the potential well distortion effect. The dots in (a,b) are obtained from turn-by-turn tracking simulations; the blue and red solid lines are evaluated by solving Eqs. (42-44) using Eqs. (40), (41), and (A5).}}
\label{Fig7}
\end{figure}

Let us first calculate the longitudinal single-bunch multi-turn dynamics based on the parameters given in Table I (the \textit{nominal} case). The numerical turn-by-turn tracking simulation and the theoretical predictions are shown in Fig.~\ref{Fig7}, where we assume the linearized synchrotron oscillation, linearized wake function and neglected the potential well distortion effect. Figures~\ref{Fig7}(a) and (b) illustrate the undulator-averaged longitudinal phase space coordinates $\bar{z}_{n}$ and $\bar{\delta}_{n}$ for the first 300 turns. The macroparticle tracking simulations are performed by numerically iterating Eqs. (\ref{Eq13},\ref{Eq14}) period by period inside the undulator and Eq.(\ref{Eq15}) turn by turn in the storage ring. The theoretical predictions are also plotted in Fig.~\ref{Fig7}(a) as a comparison. From Sec.~\ref{SecIII} the asymptotic behavior of $\bar{z}_{n}$ can be approximately written as the sum of three roots in the following form
\begin{equation}\label{Eq45}
\left| {{{\bar z}_n}} \right| \propto \sum\limits_{j = 0}^2 {{e^{ - \tau _j^{ - 1}{T_0}}}}  \approx \sum\limits_{j = 0}^2 {{e^{ - \left( {\operatorname{Im} {\Omega _{j,I}}} \right){T_0}}}}. 
\end{equation}

The red and blue curves in Fig.~\ref{Fig7}(a) are obtained based on the solved $\text{Im} \Omega_{j,I}$, in which $\hat{\mathbb{W}}$ [Eq. (\ref{Eq40})] is evaluated using the discrete sum or the continuous integration, respectively; see also Eq. (\ref{EqA5}). To better compare the theoretical predictions with the tracking simulations, an initial value on the 100-th turn is used to fit Eq. (\ref{Eq45}). Both using the continuous integration and the discrete sum give consistent predictions, and show reasonable agreement to the tracking simulation results. It deserves here to mention that the instability growth is determined by the imaginary part of the impedance, instead of the real part; see also Eq. (\ref{EqA5}). We will point out this difference in the next section when comparing such an instability with the Robinson instability. As for Fig.~\ref{Fig7}(a), the deviation between the tracking simulation and the theoretical predictions can be attributed to the underlying assumption made in Eq. (\ref{Eq18}). Figures~\ref{Fig7}(c) and (d) gives the evolution of the $z_{n}^{(p)}$ and $\delta_n^{(p)}$ at each undulator passage for different turns. The initial condition is set as $z_{0}^{(0)} = 10^{-10}$ m and $\delta_0^{(0)} = 0$. From Figs.~\ref{Fig7}(c) and (d) one can see that at the beginning both $z_{n}^{(p)}$ and $\delta_n^{(p)}$ remain uniform in the undulator. Due to the accumulated radiation wakes, $\delta_n^{(p)}$ begins to oscillate inside the undulator. It is such an oscillation inside the undulator that eventually violates the underlying assumption in Eq. (\ref{Eq18}) and leads to deviation between the theoretical predictions and the tracking simulation. Despite this, the predicted instability growth rates seem passable, considering a first order growth rate estimate.

\begin{figure}
\centering
\includegraphics[width=5in]{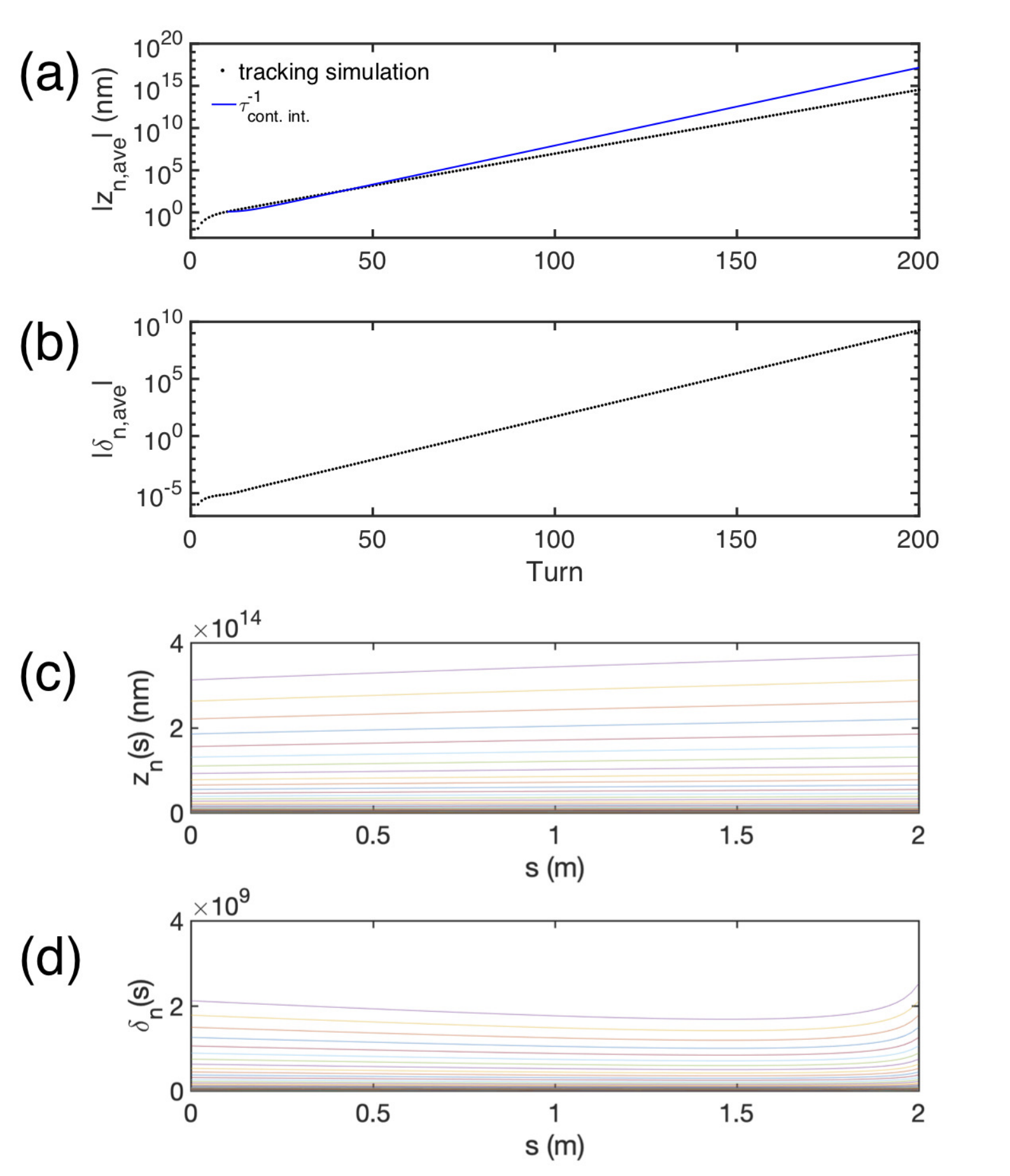}
\caption{\textmd{Evolution of the undulator-averaged phase space coordinates turn-by-turn in the storage ring (a,b) and in the undulator of the laser modulator (c,d). The beam and undulator parameters are based on Table I, except for the laser modulator voltage $V_m = 0.051$ MV. Here the linearized radiation wake function is linearized and the potential well distortion effect is neglected.}}
\label{Fig8}
\end{figure}

To further support the reason, let us reduce the longitudinal focusing by decreasing the laser modulator voltage. When the modulator voltage becomes ten times smaller and the other parameters remain the same, the synchrotron oscillation frequency is reduced by a factor of $\sqrt{10} \approx 3$. Now that Eq. (\ref{Eq18}) holds in a more satisfactory way, we expect a better matching between the theoretical prediction and the tracking simulations. Figure 8 confirms our expectation. In Figs. 8(d) one finds that the oscillation inside the undulator is not as evident as that in Fig.~\ref{Fig7}(d). Beside, the instability growth rate in this case is much larger than the nominal case in Fig.~\ref{Fig8} due to lack of the external focusing, as mentioned.

\begin{figure}
\centering
\includegraphics[width=5in]{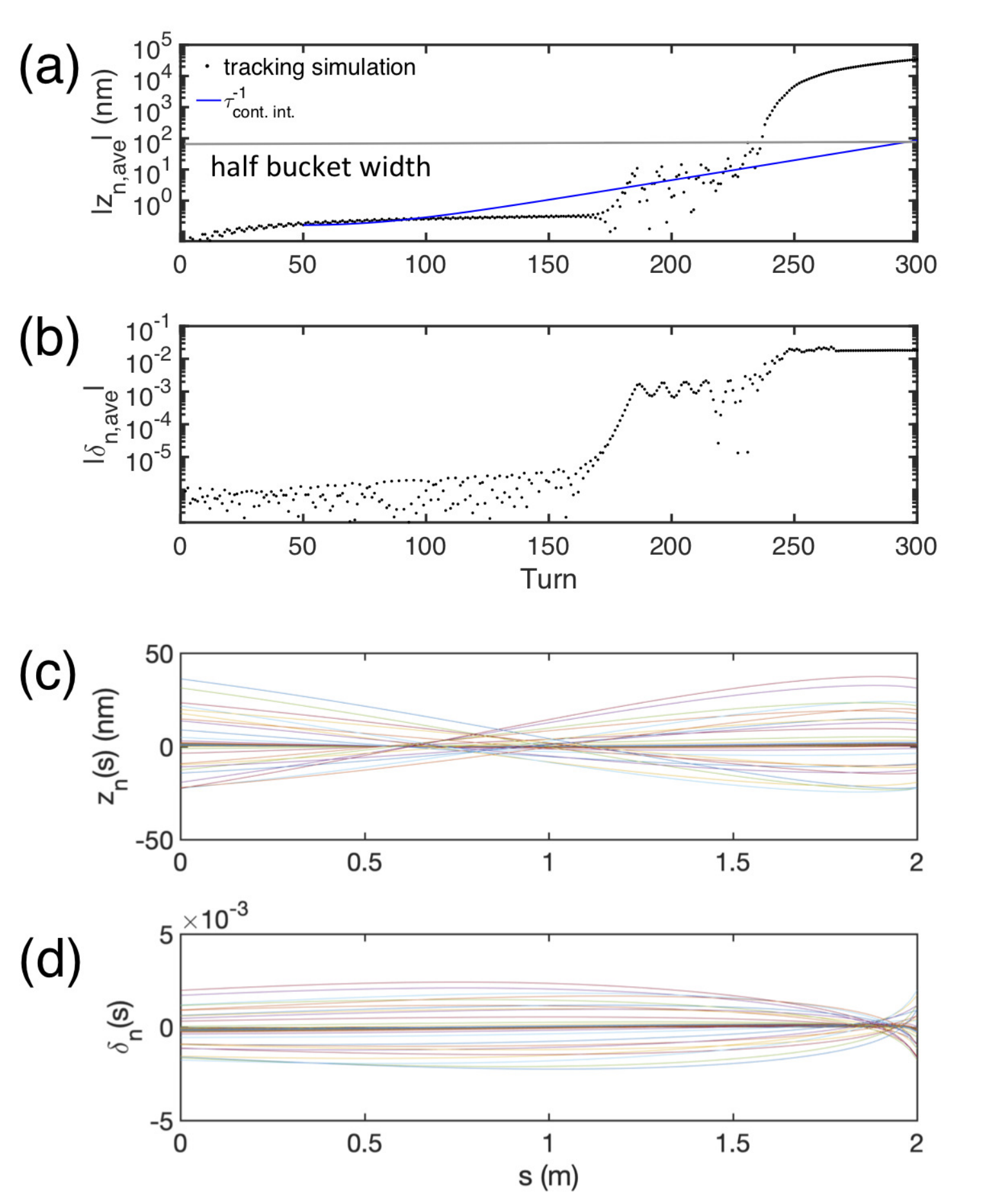}
\caption{\textmd{Evolution of the undulator-averaged phase space coordinates turn-by-turn in the storage ring (a,b) and in the undulator of the laser modulator (c,d). The beam and undulator parameters follow Table I. Here we use the original form of the radiation wake function and include the sinusoidal synchrotron motion and the potential well distortion effect. In (a) the half bucket width is indicated by the gray solid line.}}
\label{Fig9}
\end{figure}

Based on the nominal parameters given in Table I, we may wonder possible consequences if we consider the original form of the radiation wake function, which includes both the parasitic loss and potential well distortion, and retain the nonlinear, sinusoidal synchrotron oscillation. The simulation results are summarized in Fig.~\ref{Fig9}. We see from Figs.~\ref{Fig9}(a) and (b) that inclusion of the potential well effect provides additional longitudinal focusing~\cite{Ref08}, and thus the microbunch can be better confined in the phase space bucket until the first 150 turns. However starting from about the 170-th turn the bunch begins to execute large-amplitude synchrotron oscillations. After a few large-amplitude oscillations, the bunch gets lost when exceeding half the bucket width $\approx 80$ nm or half the bucket height $\delta_{\frac{1}{2}} \approx 1.7 \times 10^{-3}$. It can also be seen from Fig.~\ref{Fig9}(a) that the growth rate from the theoretical prediction is higher than that based on the tracking simulation including the aforementioned nonlinear effects.

Given the explicit expressions of the unstable solutions Eqs. (\ref{Eq42}-\ref{Eq44}) with the synchrotron frequency Eq. (\ref{Eq39}) and the scaled wake strength Eq. (\ref{Eq40}), let us take a closer look at the dependence of the storage ring slippage factor $\eta_{\text{ring}}$ on the instability growth rate. Note here that such dependence will be more involved than that presented in Fig.~\ref{Fig6} because $\omega_{s0,\text{tot}}$ and $\hat{\mathbb{W}}$ have different dependencies of $\eta_{\text{ring}}$ (or $\eta_w L_w + \eta_{\text{ring}} C_{\text{ring}}$). Figure 10 depicts the imaginary part of the three solutions Eqs. (\ref{Eq42}-\ref{Eq44}) as a function of $\eta_{\text{ring}}$. When $\eta_{\text{ring}}$ is large enough to ${\eta _w}{L_w} + {\eta _{{\text{ring}}}}{C_{{\text{ring}}}} > 0$ [see Eq. (\ref{Eq23})], the system loses single-particle optics stability, not to mention the unstable collective motion. A special situation occurs when ${\eta _w}{L_w} + {\eta _{{\text{ring}}}}{C_{{\text{ring}}}} = 0$ or ${\eta _{{\text{ring}}}} =  - {{{\eta _w}{L_w}} \mathord{\left/
 {\vphantom {{{\eta _w}{L_w}} {{C_{{\text{ring}}}}}}} \right.
 \kern-\nulldelimiterspace} {{C_{{\text{ring}}}}}} \approx 0.667 \times {10^{ - 6}}$. In this situation, $\Delta \to 0, \omega_{s0,\text{tot}} \to 0$, all the solutions result in vanishing instability growth rates $\Omega_{j,I} \to 0$. It deserves to mention that, when $\eta_{\text{ring}} < -\frac{\eta_w L_w}{C_{\text{ring}}} $, the beam is unstable at however a smaller instability growth rate.

\begin{figure}
\centering
\includegraphics[width=5in]{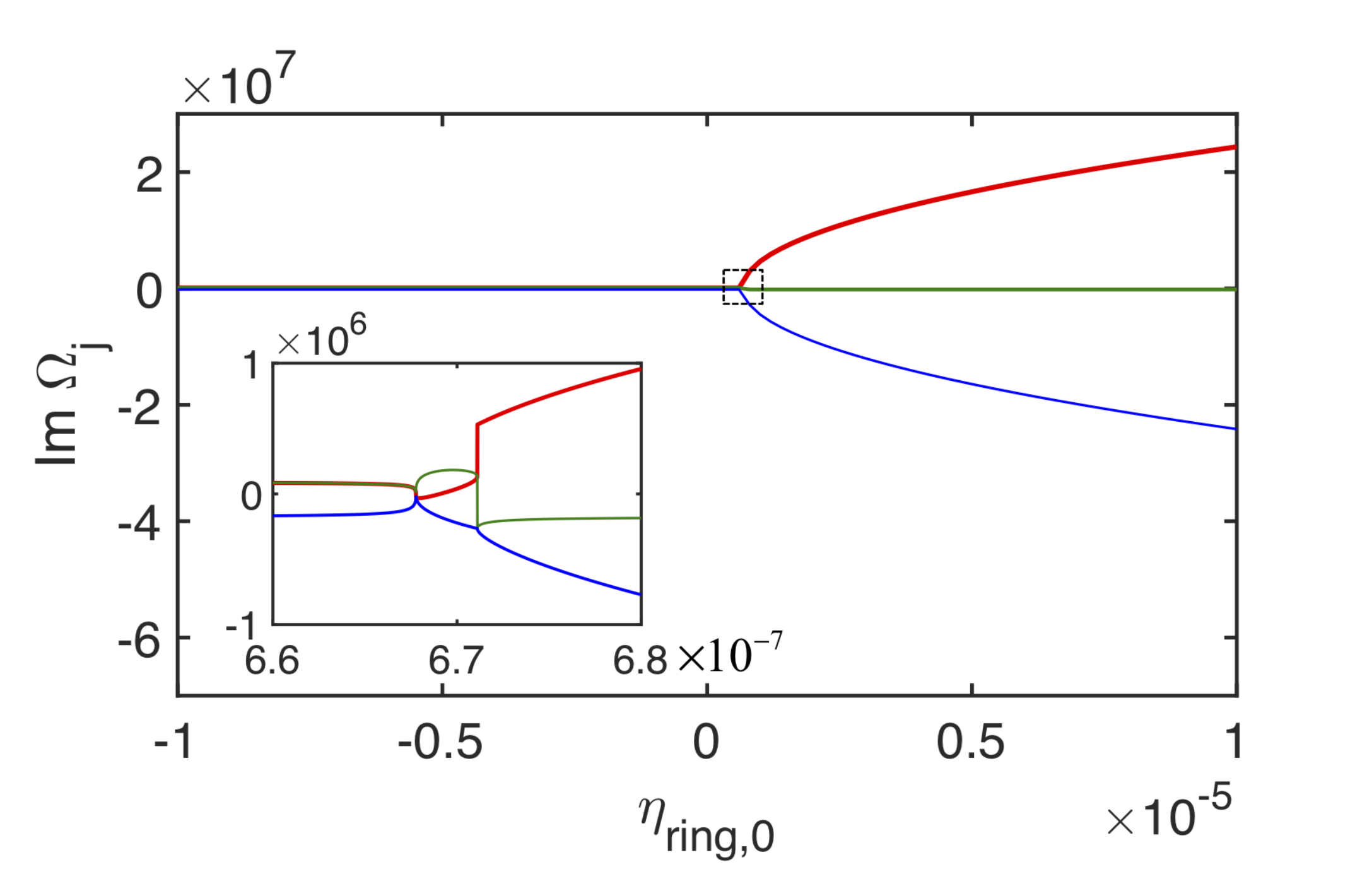}
\caption{\textmd{Dependence of the imaginary parts of the three roots in Eq. (38) [see also Eqs. (42-44)] on the storage ring slippage factor $\eta_{\text{ring,0}}$. The inset zooms in the small dashed box around $\eta_{\text{ring}} = -\eta_w L_w/C_{\text{ring}}$. The relevant parameters are given in Table I. Here $\text{Im }\Omega_j < 0$ refers to the unstable motion.}}
\label{Fig10}
\end{figure}

\begin{figure}
\centering
\includegraphics[width=5in]{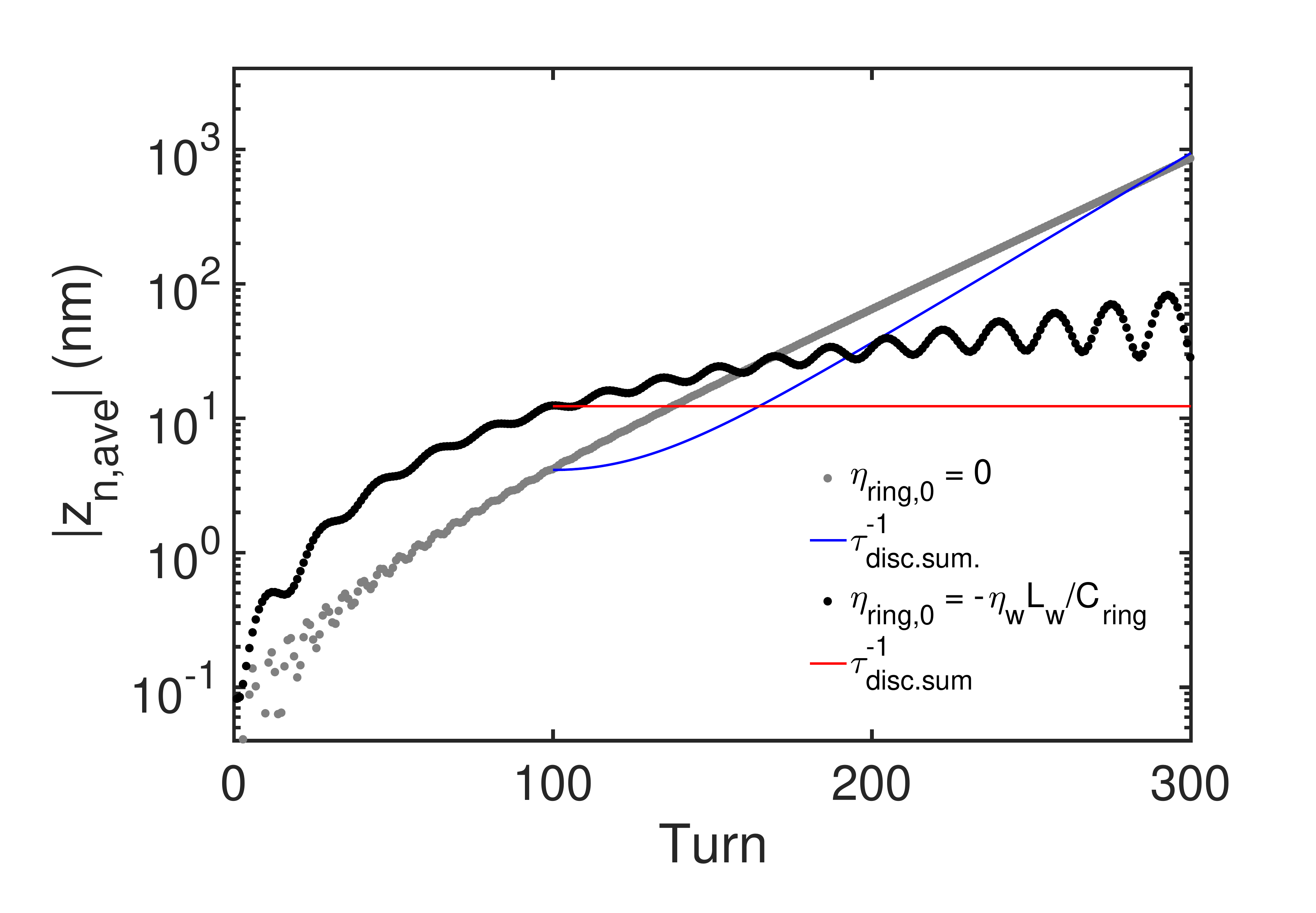}
\caption{\textmd{Evolution of the longitudinal displacements for $\eta_{\text{ring}} = 0$ and $\eta_{\text{ring}} = -\eta_w L_w/C_{\text{ring}}$.}}
\label{Fig11}
\end{figure}

Figure~\ref{Fig11} shows the turn-by-turn tracking simulations with the corresponding theoretical prediction when $\eta_{\text{ring}}$ is chosen so that $\eta_w L_w + \eta_{\text{ring}} C_{\text{ring}} = 0$. The results based on the case of $\eta_{\text{ring}} = 0$ are also illustrated in the figure for comparison [see also Fig.~\ref{Fig7}(a)]. Indeed the beam is more stable for ${\eta _{{\text{ring}}}} =  - {{{\eta _w}{L_w}} \mathord{\left/
 {\vphantom {{{\eta _w}{L_w}} {{C_{{\text{ring}}}}}}} \right.
 \kern-\nulldelimiterspace} {{C_{{\text{ring}}}}}} \approx 0.667 \times {10^{ - 6}}$ than for $\eta_{\text{ring}} = 0$; {the growth rate is zero when $\eta_w L_w + \eta_{\text{ring}} C_{\text{ring}} = 0$}. Here we again remind that, as for the deviation between the theory and the simulations, the predicted instability growth rate based on the condition ${\eta _w}{L_w} + {\eta _{{\text{ring}}}}{C_{{\text{ring}}}} = 0$ does not exactly correspond to the simulation case because we have applied the undulator-averaged phase space coordinates to simplify the analysis. 

As a side note, we may draw an analogy from Robinson instability study~\cite{Ref09,Ref10} that suppression of the instability may be possible through fine tuning between the external laser wavelength $\lambda_L$ and the radiation wavelength $\lambda_r$. Our simulation results indicate that a small detuning would not too much affect the collective dynamics. We believe that this is due to the fundamental difference of the wake functions between such an instability and the Robinson one. In the next section we will make a comparison between the two instability mechanisms. An option might be possible to detune the path length of the circulating microbunch to mitigate such instability. This option relies upon a change of the storage ring slippage factor $\eta_{\text{ring}}$, as we have illustrated in Fig.~\ref{Fig10}.

Before ending this section we point out that the number of electrons in the microbunch does not only determine the instability but also is determined by other effects, particularly the output performance of the undulator in the radiator and those during revolution in the storage ring outside the laser cavity modulator. As for possible manipulation utilizing the effects of the cavity mirrors on the accumulated radiation fields, a dedicated cavity system design with proper mirror reflectivity may be desired. Further studies on this topic have been beyond the scope of this paper.

\section{Discussion}\label{SecV}
In this paper we have developed a theoretical formulation for the coherent radiation induced multi-turn instability in a storage ring including the laser cavity modulator. In this section we make a comparison between such an instability and the Robinson one. The Robinson instability that we will discuss below refers to the single-bunch multi-turn instability driven by the RF cavity induced long-range wakefield. First, for Robinson case, the wake function that a circulating electron sees on the $n$-th turn from the $k$-th turn can be written as~\cite{Ref09,Ref10}
\begin{equation}\label{Eq46}
{W_\parallel }\left( {kC - nC + {z_n} - {z_k}} \right) \approx {W_\parallel }\left( {kC - nC} \right) + {W_\parallel^{\prime} }\left( {kC - nC} \right)\left( {{z_n} - {z_k}} \right)
\end{equation}
where $C$ is the storage ring circumference and $z_{n,k}$ are the longitudinal displacement of the bunch on the $n$-th and $k$-th turn, respectively. Note that the convention $W_{\parallel} (z > 0) = 0$ is given. An approximate expression in Eq. (\ref{Eq46}) can be made by Taylor expanding the argument, considering the fact that $|k-n|C \gg |z_n-z_k|$. Similar to the first term in Eq. (\ref{Eq17}), the first term on the right hand side characterizes the parasitic loss of the entire bunch. The sum of Eq. (\ref{Eq46}) over the previous revolutions $k$, $\sum\limits_{k = 0}^{n - 1} {{{W}_\parallel^{\prime} }(kC - nC)\left( {{z_n} - {z_k}} \right)} $, is of the convolution type. In contrast, the sum Eq. (\ref{Eq19}) over the undulator period $p$, or Eq. (\ref{EqA1}), is of non-convolution type. The reason for such a difference lies in the fact that, for the Robinson case, the wake function is located in the RF cavity and the circulating bunch consecutively samples the temporally decaying wakefield in a regular separation equal to the storage ring circumference. For our case, the radiation wake propagates in the undulator of the laser modulator and the circulating bunch synchronously meets the radiation fields in a phase-locking manner. The radiation wakes, sampled consecutively by the traversing electron bunch along the undulator, are with the finite duration of $N_w\lambda_r$.

Based on the macroparticle equations of motion with the functional form of Eq. (\ref{Eq46}), the resultant dispersion equation is quadratic~\cite{Ref09,Ref10} and gives two solutions: one for the exponential growth and the other for the exponential damping. Through a proper detuning between the RF frequency and the harmonic revolution frequency can damp Robinson instability in a conventional storage ring. In contrast, at least one of the three solutions based on the cubic dispersion equation Eq. (\ref{Eq38}) [or Eqs. (\ref{Eq40}-\ref{Eq42})] remains unstable{, except for the situation when $\eta_w L_w + \eta_{\text{ring}}C_{\text{ring}} = 0$. It has been known that in a conventional storage ring, a proper and clean detuning of the RF cavity to suppress Robinson instability works because the cavity, where the impedances reside, does not have a finite $R_{56}$ (i.e., $\eta_{\text{RF}} = 0$). That is, such RF cavity does not change the longitudinal bunch centroid on its passage, acting as a longitudinal \textit{thin}-lens element. The situation is different in the laser modulator cavity. The effect of $R_{56,w} = \eta_w L_w$ is equivalent to treating an undulator-impedance element as a longitudinal \textit{thick}-lens element. Inside the modulator undulator the electrons execute a wiggling motion, and the undulator has non-vanishing $R_{56,w}$. Although Robinson-like instability mechanism can still be present in the modulator cavity, its detailed mechanism has changed. The non-vanishing $R_{56,w}$ inside the modulator cavity will smear out the longitudinal motion, thus erasing the clean Robinson mechanism. Therefore we may expect that the clean criterion to suppress the instability becomes less pronounced. Notice that the modulator undulator $R_{56,w}$ can be comparable to the storage ring $R_{56,\text{ring}} = \eta_{\text{ring}}C_{\text{ring}}$. Therefore the effect of $R_{56,w}$ or/and $R_{56,\text{ring}}$ can be sensitive, especially for the very short SSMB microbunch.} {From the macroparticle model, it appears unlikely to} eliminate the instability without any external damping mechanism. To some extent, the two aforementioned options can only mitigate the instability (Fig.~\ref{Fig10}). Moreover, the instability growth rate of the Robinson type is attributed to the real part of the impedance, while the instability growth rate of our case is related to the imaginary part of the radiation impedance.

The cubic dispersion equation may be related to the FEL instability as a special case. We have drawn a similar analogy in Ref.~\cite{Ref08} between the coherent radiation driven longitudinal single-pass beam breakup instability and the post-saturation FEL sideband instability (see, e.g., Refs~\cite{Ref39,Ref40,Ref41}). Here we remind that a more in-depth and quantitative discussion, including possible connection to the FEL instability~\cite{Ref32}, will be published in a separate paper.

{Before ending this section, we add some comments on the wake-impedance approach used in the analysis, the validity of using macroparticle model, and some different features of the laser modulator cavity from the conventional RF cavity. First, using the wake-impedance approach may not be a very ideal way for the analysis. The classic situation where the wake-impedance approach is employed assumes the particle beam traverses a straight trajectory, and the wake field is spatially localized~\cite{Ref09}. For our case, the particle beam executes a wiggling motion in the modulator undulator, and the radiation field travels with the beam and is no longer localized. The use of the wake-impedance approach has therefore some limitations. For the non-straight, wiggling motion, if we define the wake or impedance by relating the energy transfer (i.e., work done) through the transverse component of the radiation field coupled with the wiggling motion of the particle, the dynamical process can still be formulated in the 1-D framework. This has been widely applied in the 1-D FEL analysis and proved successful in exploring many FEL essences. As for the traveling wake field, if we properly take into consideration the slippage between the particle and the radiation field (due to the velocity difference in the undulator), the wake-impedance approach can still work in formulating the FEL or FEL-like process. This part was discussed in Ref.~\cite{Ref44}. A more self-consistent approach describing the FEL process requires explicit equations of motion for the particle phase-space coordinates and the radiation field (see for example Ref.~\cite{Ref01}). In the wake-impedance description, only the particle's equations of motion are explicit; the radiation field dynamics has been implicitly embedded in the formulation. While we believe that the self-consistent approach is more complete, the analysis based on the wake-impedance approach can be simpler in our situation (the multi-turn single microbunch dynamics). The next step will be to re-formulate this problem using the self-consistent approach and re-examine the microbunch dynamics.

Second, in the analysis the electron beam in the laser modulator has already been microbunched, and the total bunch length is assumed much shorter than the external laser wavelength. Our starting point assumes that the microbunch is well developed in the laser modulator of an SSMB storage ring. This is similar to the situation where an electron bunch is well developed in the RF wavelength of a conventional storage ring. The collective dynamics of our interest is studied based on this starting point. The transition from the unbunched beam to the microbunched beam, if not externally injected, is not of our current interest in this paper. The radiation wake generated by the microbunch is thus coherent because the already well developed microbunch only occupies a small portion of the phase space bucket. Thus bunching is assumed to always be at a maximum in our macroparticle description. To be more specific, the bunching here refers to the modulation laser wavelength, instead of the SSMB radiation wavelength. When using a macroparticle to represent a microbunch, we expect that the microbunch length is much shorter than the laser wavelength. In the laser modulator we care more about the multi-turn dynamics of the bulk microbunch. Together with the fact of the microbunch length being much shorter than the laser wavelength, the bunching factor is not a dynamical variable here. Now that a microbunch is represented by a macroparticle, when the centroid oscillation amplitude is too large, we expect that the bunching factor or beam quality will degrade. If a more detailed study of the microbunch dynamics is needed, the bunching factor should be considered as a dynamic quantity. When studying the beam-radiation dynamics in the radiator undulator, the bunching factor, now with respect to the SSMB desired radiation wavelength, should be treated as a dynamic quantity. 

Third, it is known that the RF cavities are mostly single-moded, and the optical cavities are highly multi-moded, both longitudinally and transversely. In the laser modulator cavity, according to the estimate in Refs.~\cite{Ref30,Ref31}, the linewidth should be within a few kHz up to MHz. This number is much smaller than the typical spectral width of the modulator undulator spectrum. Compared with the radiation spectral width about $1/2N_w$, the linewidth is negligibly small. Thus we believe that, within the time scale of the interested dynamical process (the instability growth), the radiation field in the laser modulator cavity can be derived by the radiation field in the free space, as we have outlined in Section II.A. Another important point is the presence of the cavity mirrors. The effects of the cavity mirrors typically include the field dissipation and the frequency filtering. These are important effects to the microbunch dynamics; however we leave these comprehensive investigations in our future work, primarily because the cavity mirror design depends on the choice of the material, the geometry and other laser-optics aspects. To focus on the multi-turn instability mechanism in this paper, we assume that the cavity mirrors are ideal devices, i.e., no field dissipation and the reflectance is wavelength-independent.

In the presented numerical example, we only consider the most dominant, fundamental harmonic of the modulator undulator radiation. We believe that this simplification is reasonable; the higher harmonics of the radiation fields, after averaging over the undulator periods, may have negligible effect on the beam, at least on the bulk of the microbunch. This simplification is also consistent with the macroparticle model used in the analysis. The higher harmonics of the radiation field on the microbunch may be reflected on the internal phase space structure of the microbunch. This underlying physics is however excluded from the single macroparticle model and thus beyond the scope of this work.}

\section{Summary and Outlook}\label{SecVI}
In this paper we have formulated the longitudinal single-bunch multi-turn collective dynamics in a storage ring with a laser modulator cavity. Based on the macroparticle model, the equations of motion for the microbunch in the undulator and in the remaining storage ring have been constructed as two sets of difference equations and can be solved by using z-transform. Then we obtained the simplified, analytical formulae [Eqs. (\ref{Eq42}-\ref{Eq44})] of the instability growth rate by introduction of the undulator-averaged phase space coordinates. Compared with the Robinson instability, an instability driven by the long-range RF cavity induced wake, the essential difference lies in the fact that in our case the undulator radiation wake is no longer localized at a location, but propagating inside the cavity modulator with a finite duration. The circulating electron bunch then receives the radiation energy kick at the undulator entrance in the phase-locking manner with the external laser in the modulator. Such a characteristic leads to a different form of the dispersion equation from the Robinson case. The developed formulation is then applied to a set of preliminary design parameters of the SSMB storage ring to illustrate the coherent undulator radiation induced longitudinal single-bunch collective dynamics. The theoretical predictions of the linear instability growth rate show good agreement with the turn-by-turn tracking simulations. Based on the parameters given in Table I, our calculations indicate that, in the absence of any damping mechanism, this single microbunch may remain stable for the first hundred revolutions operating in the single-particle optics stability regime. After about 150 revolutions, due to the nonlinear synchrotron motion and the potential well distortion effect, the bunch centroid starts large-amplitude oscillations and get lost when the oscillation amplitude exceeds the half-bucket width. {Here we remark that in the numerical example the assumed bunch charge 6.4 fC is about one order of magnitude larger than the value assumed in the most recent SSMB designs, i.e., 0.64 fC with the average microbunch current of 1 A. The instability growth rate correspondingly decreases with one order of magnitude. However from the macroparticle analysis the system appears to be essentially unstable provided no damping mechanism.}

Here we point out that, although using the macoparticle model is far from a complete description of the microbunch dynamics, we place emphasis on such a potential instability mechanism in the laser modulator cavity of the SSMB storage ring. A more dedicated SSMB storage ring design to generate and maintain a short bunch may require implementation of a quasi-isochronous lattice with strong longitudinal focusing~\cite{Ref42}. In that situation a more complete analysis may include the impact of higher order momentum compaction factors on the instability, including the storage ring lattice and the undulators from both the cavity modulator and the radiator [Eq. (\ref{Eq16})]. In the meanwhile other relevant physical effects from, e.g., cavity mirrors and radiator effects should also be incorporated. As for the single-bunch beam dynamics, the present analysis based on the macroparticle model only reflects the dipole motion. Those mitigating effects arising from the finite energy spread and transverse beam emittances are not yet considered. A more complete analysis should take these physical effects into account. Besides, a practical extension of the analysis to the multi-bunch multi-turn collective dynamics is ongoing.

\begin{acknowledgements}
The author would like to express sincere gratitude to Prof. Alexander Wu Chao (Stanford), Dr. Yi Jiao (IHEP), Dr. Xiujie Deng, Zizheng Li, Dr. Zhilong Pan, Prof. Chuanxiang Tang, Dr. Yao Zhang, and Dr. Kaishang Zhou (THU) for many stimulating and insightful discussions on this work and SSMB design concepts and Drs. Qinghong Zhou (SUST), Hao-Wen Luo (NTHU) and Mr. Make Ying for many helpful discussions and comments on this manuscript. He also acknowledges W. Tsai for help with the preparation of this manuscript. This work is supported by the Fundamental Research Funds for the Central Universities under Project No. 5003131049 and National Natural Science Foundation of China under project No. 11905073.
\end{acknowledgements}

\appendix
\section{Detailed derivation of Eq. (21)}
This appendix contains the derivation that completes Eq. (\ref{Eq21}) from Eq. (\ref{Eq14}). We first Taylor expand and linearize the radiation wake functions
\begin{equation}\label{EqA1}
\sum\limits_{p = 0}^{{N_w} - 1} {{{\mathcal{W}}_\parallel }\left[ {\left( {{N_w} - p} \right){\lambda _r} + z_m^{(p)} - z_k^{(0)}} \right]}  \approx \sum\limits_{p = 0}^{{N_w} - 1} {\left\{ {{{\mathcal{W}}_\parallel }\left[ {\left( {{N_w} - p} \right){\lambda _r}} \right] + \left( {z_m^{(p)} - z_k^{(0)}} \right){{{\mathcal{W}}}_\parallel^{\prime} }\left[ {\left( {{N_w} - p} \right){\lambda _r}} \right]} \right\}}.
\end{equation}
Neglecting the parasitic loss term and the potential distortion term, we are left with the following sum
\begin{align}\label{EqA2}
{\mathbb{W}'} &= \sum\limits_{p = 0}^{{N_w} - 1} {{{{\mathcal{W}}}_\parallel^{\prime} }\left[ {\left( {{N_w} - p} \right){\lambda _r}} \right]}  = \frac{{4\pi {\epsilon _0}N{r_e}{\lambda _w}}}{{{\gamma _0}}}\sum\limits_{p = 0}^{{N_w} - 1} {{{W}_\parallel^{\prime} }\left[ {\left( {{N_w} - p} \right){\lambda _r}} \right]}  \nonumber \\
&= \frac{{4\pi {\epsilon _0}N{r_e}{\lambda _w}}}{{{\gamma _0}}}\frac{i}{{2\pi c}}\int_{ - \infty }^\infty  {\omega {Z_\parallel }(\omega )\left[ {\sum\limits_{p = 0}^{{N_w} - 1} {{e^{i\frac{\omega }{c}\left( {{N_w} - p} \right){\lambda _r}}}} } \right]{\text{d}}\omega }, 
\end{align}
where the last equality is obtained by expressing the derivative of the wake function in terms of the corresponding impedance function~\cite{Ref09}. The discrete sum in the square bracket can be analytically evaluated
\begin{equation}\label{EqA3}
\sum\limits_{p = 0}^{{N_w} - 1} {{e^{i\frac{\omega }{c}\left( {{N_w} - p} \right){\lambda _r}}}}  = {e^{i\frac{\omega }{c}\left( {\frac{{{N_w} + 1}}{2}} \right){\lambda _r}}}\frac{{\sin \tfrac{{{N_w}}}{2}\tfrac{\omega }{c}{\lambda _r}}}{{\sin \tfrac{1}{2}\tfrac{\omega }{c}{\lambda _r}}} = {e^{i2\pi \tilde k\left( {\frac{{{N_w} + 1}}{2}} \right)}}\frac{{\sin {N_w}\pi \tilde k}}{{\sin \pi \tilde k}},
\end{equation}
where we define $\tilde k = {k \mathord{\left/ {\vphantom {k {{k_r}}}} \right. \kern-\nulldelimiterspace} {{k_r}}}$. The ratio $\sin N_w\pi\tilde{k}/\sin \pi \tilde{k}$ is in fact the Chebyshev polynomial of the second kind~\cite{Ref43} and takes an appreciable value around the integers in the argument. Taylor expanding this function, we have
\begin{align}\label{EqA4}
\frac{{\sin {N_w}\pi \tilde k}}{{{N_w}\sin \pi \tilde k}} = {U_{{N_w} - 1}}\left( {\cos \pi \tilde k} \right) &= \left\{ \begin{gathered}
  \sum\limits_n^{} {{{\left( { - 1} \right)}^{{N_w}}}\left[ { - 1 + \frac{{{\pi ^2}}}{6}N_w^2{{\left( {\tilde k - n} \right)}^2} -  +  \cdots } \right]} {\text{,    for odd }}n \hfill \\
  \sum\limits_n^{} {\left( { - 1} \right)\left[ { - 1 + \frac{{{\pi ^2}}}{6}N_w^2{{\left( {\tilde k - n} \right)}^2} -  +  \cdots } \right]} {\text{,       for even }}n \hfill \\ 
\end{gathered}  \right. \nonumber \\
& \approx \sum\limits_n^{} {{{\left( { - 1} \right)}^{{N_w}n}}\delta \left( {\tilde k - n} \right)},
\end{align}
which indicates that the first order approximate of this expression can be written as a sum of Kronecker delta functions. Thus Eq. (\ref{EqA2}) can be further simplified in terms of the normalized wavenumber
\begin{align}\label{EqA5}
{\mathbb{W}'} &= \frac{{4\pi {\epsilon _0}N{r_e}{\lambda _w}}}{{{\gamma _0}}}\frac{{ik_r^2c{N_w}}}{{2\pi }}\int_{ - \infty }^\infty  {\tilde k{Z_\parallel }(\tilde k)\left[ {{e^{i2\pi \tilde k\left( {\frac{{{N_w} + 1}}{2}} \right)}}\frac{{\sin {N_w}\pi \tilde k}}{{{N_w}\sin \pi \tilde k}}} \right]{\text{d}}\tilde k}  \nonumber \\
& \approx \frac{{4\pi {\epsilon _0}N{r_e}{\lambda _w}}}{{{\gamma _0}}}\frac{{ik_r^2c{N_w}}}{{2\pi }}\sum\limits_{n =  - \infty }^\infty  {{{\left( { - 1} \right)}^{{N_w}n}}n{Z_\parallel }(n)}   \nonumber \\
& =  - \frac{{4\pi {\epsilon _0}N{r_e}{\lambda _w}}}{{{\gamma _0}}}\frac{{k_r^2c{N_w}}}{{2\pi }}\sum\limits_{n =  - \infty }^\infty  {{{\left( { - 1} \right)}^{{N_w}n}}n\operatorname{Im} {Z_\parallel }(n)}, 
\end{align}
where the approximation $N_w \gg 1$ is made in the second line. Thus the continuous integration can be replaced by the discrete sum, i.e., $\int {\left( {...} \right){\text{d}}\tilde k}  \to \sum\limits_{n =  - \infty }^\infty  {\left( {...} \right)\delta \left( {\tilde k - n} \right)}$. In our case with $N_w = 100$, the result from the continuous integration does not make much difference from that of discrete sum. The last equality in Eq. (\ref{EqA5}) is based on the symmetry properties of the impedance with
\begin{align}\label{EqA6}
i\sum\limits_{n =  - \infty }^\infty  {{{\left( { - 1} \right)}^{{N_w}n}}n{Z_\parallel }(n)} &= i\sum\limits_{n =  - \infty }^\infty  {{{\left( { - 1} \right)}^{{N_w}n}}n\left[ {\operatorname{Re} {Z_\parallel }(n) + i\operatorname{Im} {Z_\parallel }(n)} \right]}  \nonumber \\
&=  - \sum\limits_{n =  - \infty }^\infty  {{{\left( { - 1} \right)}^{{N_w}n}}n\operatorname{Im} {Z_\parallel }(n)}. 
\end{align}
Here we comment on our evaluation of the imaginary part of the radiation impedance. We first obtain the radiation wake function $W_{\parallel}$ from Eqs. (\ref{Eq1}) and (\ref{Eq3}). The corresponding radiation impedance can be retrieved from Fourier transformation of the wake function. The real part of the retrieved impedance is the same as the Eq. (\ref{Eq3}). The imaginary part is then used in Eq. (\ref{EqA5}).



\begin{thebibliography}{0}%
\makeatletter
\providecommand \@ifxundefined [1]{%
 \@ifx{#1\undefined}
}%
\providecommand \@ifnum [1]{%
 \ifnum #1\expandafter \@firstoftwo
 \else \expandafter \@secondoftwo
 \fi
}%
\providecommand \@ifx [1]{%
 \ifx #1\expandafter \@firstoftwo
 \else \expandafter \@secondoftwo
 \fi
}%
\providecommand \natexlab [1]{#1}%
\providecommand \enquote  [1]{``#1''}%
\providecommand \bibnamefont  [1]{#1}%
\providecommand \bibfnamefont [1]{#1}%
\providecommand \citenamefont [1]{#1}%
\providecommand \href@noop [0]{\@secondoftwo}%
\providecommand \href [0]{\begingroup \@sanitize@url \@href}%
\providecommand \@href[1]{\@@startlink{#1}\@@href}%
\providecommand \@@href[1]{\endgroup#1\@@endlink}%
\providecommand \@sanitize@url [0]{\catcode `\\12\catcode `\$12\catcode
  `\&12\catcode `\#12\catcode `\^12\catcode `\_12\catcode `\%12\relax}%
\providecommand \@@startlink[1]{}%
\providecommand \@@endlink[0]{}%
\providecommand \url  [0]{\begingroup\@sanitize@url \@url }%
\providecommand \@url [1]{\endgroup\@href {#1}{\urlprefix }}%
\providecommand \urlprefix  [0]{URL }%
\providecommand \Eprint [0]{\href }%
\providecommand \doibase [0]{https://doi.org/}%
\providecommand \selectlanguage [0]{\@gobble}%
\providecommand \bibinfo  [0]{\@secondoftwo}%
\providecommand \bibfield  [0]{\@secondoftwo}%
\providecommand \translation [1]{[#1]}%
\providecommand \BibitemOpen [0]{}%
\providecommand \bibitemStop [0]{}%
\providecommand \bibitemNoStop [0]{.\EOS\space}%
\providecommand \EOS [0]{\spacefactor3000\relax}%
\providecommand \BibitemShut  [1]{\csname bibitem#1\endcsname}%
\let\auto@bib@innerbib\@empty
\end{thebibliography}%


\begin{thebibliography}{10}

\bibitem{Ref01}
Kwang-Je Kim, Zhirong Huang, and Ryan Lindberg.
\newblock {\em Synchrotron radiation and free-electron lasers}.
\newblock Cambridge university press, 2017.

\bibitem{Ref02}
Daniel~F. Ratner and Alexander~W. Chao.
\newblock Steady-state microbunching in a storage ring for generating coherent
  radiation.
\newblock {\em Physical review letters}, 105(15):154801, 2010.

\bibitem{Ref03}
Yi~Jiao, Daniel~F. Ratner, and Alexander~W. Chao.
\newblock Terahertz coherent radiation from steady-state microbunching in
  storage rings with x-band radio-frequency system.
\newblock {\em Physical Review Special Topics-Accelerators and Beams},
  14(11):110702, 2011.

\bibitem{Ref04}
A.~Chao, E.~Granados, X.B. Huang, D.~Ratner, and H.-W. Luo.
\newblock High power radiation sources using the steady-state microbunching
  mechanism.
\newblock {\em 7th International Particle Accelerator Conference, Busan, Korea
  IPAC16 (TUXB01)}, 2016.

\bibitem{Ref05}
Chuanxiang Tang, Xiujie Deng, Wenhui Huang, Tenghui Rui, Alex Chao, J{\"o}rg
  Feikes, Ji~Li, Markus Ries, Chao Feng, Bocheng Jiang, et~al.
\newblock An overview of the progress on {SSMB}.
\newblock In {\em the 60th ICFA Advanced Beam Dynamics Workshop on Future Light
  Sources, Shanghai}, 2018.

\bibitem{Ref06}
X.J. Deng, A.W. Chao, J.~Feikes, W.H. Huang, M.~Ries, and C.X. Tang.
\newblock Single-particle dynamics of microbunching.
\newblock {\em Physical Review Accelerators and Beams}, 23(4):044002, 2020.

\bibitem{Ref07}
Xiujie Deng, Alexander Chao, J{\"o}rg Feikes, Arne Hoehl, Wenhui Huang, Roman
  Klein, Arnold Kruschinski, Ji~Li, Aleksandr Matveenko, Yuriy Petenev, et~al.
\newblock Experimental demonstration of the mechanism of steady-state
  microbunching.
\newblock {\em Nature}, 590(7847):576--579, 2021.

\bibitem{Ref08}
Cheng-Ying Tsai, Alexander~Wu Chao, Yi~Jiao, Hao-Wen Luo, Make Ying, and
  Qinghong Zhou.
\newblock Coherent-radiation-induced longitudinal single-pass beam breakup
  instability of a steady-state microbunch train in an undulator.
\newblock {\em Phys. Rev. Accel. Beams}, 24:114401, Nov 2021.

\bibitem{Ref09}
Alexander~Wu Chao.
\newblock {\em Physics of collective beam instabilities in high energy
  accelerators}.
\newblock John Wiley \& Sons, 1993.

\bibitem{Ref10}
King-Yuen Ng.
\newblock {\em Physics of intensity dependent beam instabilities}.
\newblock World Scientific, 2006.

\bibitem{Ref11}
E.L. Saldin, E.A. Schneidmiller, and M.V. Yurkov.
\newblock Klystron instability of a relativistic electron beam in a bunch
  compressor.
\newblock {\em Nuclear Instruments and Methods in Physics Research Section A:
  Accelerators, Spectrometers, Detectors and Associated Equipment},
  490(1-2):1--8, 2002.

\bibitem{Ref12}
S.~Heifets, G.~Stupakov, and S.~Krinsky.
\newblock Coherent synchrotron radiation instability in a bunch compressor.
\newblock {\em Physical Review Special Topics-Accelerators and Beams},
  5(6):064401, 2002.

\bibitem{Ref13}
Zhirong Huang and Kwang-Je Kim.
\newblock Formulas for coherent synchrotron radiation microbunching in a bunch
  compressor chicane.
\newblock {\em Physical Review Special Topics-Accelerators and Beams},
  5(7):074401, 2002.

\bibitem{Ref14}
Marco Venturini.
\newblock Microbunching instability in single-pass systems using a direct
  two-dimensional {V}lasov solver.
\newblock {\em Physical Review Special Topics-Accelerators and Beams},
  10(10):104401, 2007.

\bibitem{Ref15}
Marco Venturini, Robert Warnock, and Alexander Zholents.
\newblock Vlasov solver for longitudinal dynamics in beam delivery systems for
  x-ray free electron lasers.
\newblock {\em Physical Review Special Topics-Accelerators and Beams},
  10(5):054403, 2007.

\bibitem{Ref16}
Cheng-Ying Tsai, Weilun Qin, Kuanjun Fan, Xiaofan Wang, Juhao Wu, and Guanqun
  Zhou.
\newblock Theoretical formulation of phase space microbunching instability in
  the presence of intrabeam scattering for single-pass or recirculation
  accelerators.
\newblock {\em Physical Review Accelerators and Beams}, 23(12):124401, 2020.

\bibitem{Ref17}
Cheng-Ying Tsai and Weilun Qin.
\newblock Semi-analytical analysis of high-brightness microbunched beam
  dynamics with collective and intrabeam scattering effects.
\newblock {\em Physics of Plasmas}, 28(1):013112, 2021.

\bibitem{Ref18}
C.-Y. Tsai and W.~Qin.
\newblock {Microbunching Instability in the Presence of Intrabeam Scattering
  for Single-Pass Accelerators}.
\newblock In {\em Proc. IPAC'21}, number~12 in International Particle
  Accelerator Conference, pages 3692--3695. JACoW Publishing, Geneva,
  Switzerland, 08 2021.
\newblock https://doi.org/10.18429/JACoW-IPAC2021-THXA04.

\bibitem{Ref19}
Marco Venturini, Robert Warnock, Ronald Ruth, and James~A. Ellison.
\newblock Coherent synchrotron radiation and bunch stability in a compact
  storage ring.
\newblock {\em Physical Review Special Topics-Accelerators and Beams},
  8(1):014202, 2005.

\bibitem{Ref20}
G.~Stupakov and S.~Heifets.
\newblock Beam instability and microbunching due to coherent synchrotron
  radiation.
\newblock {\em Physical Review Special Topics-Accelerators and Beams},
  5(5):054402, 2002.

\bibitem{Ref21}
S.~Heifets and G.~Stupakov.
\newblock Single-mode coherent synchrotron radiation instability.
\newblock {\em Physical Review Special Topics-Accelerators and Beams},
  6(6):064401, 2003.

\bibitem{Ref22}
S.~Heifets.
\newblock Single-mode coherent synchrotron radiation instability of a bunched
  beam.
\newblock {\em Physical Review Special Topics-Accelerators and Beams},
  6(8):080701, 2003.

\bibitem{Ref23}
Yunhai Cai.
\newblock Linear theory of microwave instability in electron storage rings.
\newblock {\em Physical Review Special Topics-Accelerators and Beams},
  14(6):061002, 2011.

\bibitem{Ref24}
C.-Y. Tsai, D.~Douglas, R.~Li, and C.~Tennant.
\newblock Linear microbunching analysis for recirculation machines.
\newblock {\em Physical Review Accelerators and Beams}, 19(11):114401, 2016.

\bibitem{Ref25}
C.-Y. Tsai, Ya.~S. Derbenev, D.~Douglas, R.~Li, and C.~Tennant.
\newblock Vlasov analysis of microbunching instability for magnetized beams.
\newblock {\em Physical Review Accelerators and Beams}, 20(5):054401, 2017.

\bibitem{Ref26}
Cheng-Ying Tsai, Simone Di~Mitri, David Douglas, Rui Li, and Chris Tennant.
\newblock Conditions for coherent-synchrotron-radiation-induced microbunching
  suppression in multibend beam transport or recirculation arcs.
\newblock {\em Physical Review Accelerators and Beams}, 20(2):024401, 2017.

\bibitem{Ref27}
Cheng-Ying Tsai.
\newblock Concatenated analyses of phase space microbunching in high brightness
  electron beam transport.
\newblock {\em Nuclear Instruments and Methods in Physics Research Section A:
  Accelerators, Spectrometers, Detectors and Associated Equipment},
  940:462--474, 2019.

\bibitem{Ref28}
Cheng-Ying Tsai.
\newblock An alternative view of coherent synchrotron radiation induced
  microbunching development in multibend recirculation arcs.
\newblock {\em Nuclear Instruments and Methods in Physics Research Section A:
  Accelerators, Spectrometers, Detectors and Associated Equipment}, 943:162499,
  2019.

\bibitem{Ref29}
X.J. Deng, R.~Klein, A.W. Chao, A.~Hoehl, W.H. Huang, J.~Li, J.~Lubeck,
  Y.~Petenev, M.~Ries, I.~Seiler, et~al.
\newblock Widening and distortion of the particle energy distribution by
  chromaticity in quasi-isochronous rings.
\newblock {\em Physical Review Accelerators and Beams}, 23(4):044001, 2020.

\bibitem{Ref30}
Qinghong Zhou, Lisha Tu, Liuying Chen, Jun Liu, Ma-Ke Ying, Karen Lei, Hao-Wen
  Luo, Juhao Wu, and Alexander~W Chao.
\newblock Nanosecond pulse enhancement in narrow linewidth cavity for
  steady-state microbunching.
\newblock In {\em High Intensity Lasers and High Field Phenomena}, pages
  JM3A--5. Optical Society of America, 2020.

\bibitem{Ref31}
Q.H. Zhou.
\newblock {Nanosecond Pulse Enhancement in Narrow Linewidth Cavity for
  Steady-State Microbunching}.
\newblock In {\em Proc. FEL'19}, number~39 in Free Electron Laser Conference,
  pages 697--699. JACoW Publishing, Geneva, Switzerland, nov 2019.
\newblock https://doi.org/10.18429/JACoW-FEL2019-THP054.

\bibitem{Ref33}
Alexander~Wu Chao.
\newblock {\em Lectures on Accelerator Physics}.
\newblock World Scientific, 2020.

\bibitem{Ref35}
Helmut Wiedemann.
\newblock {\em Particle accelerator physics, 4th ed.}
\newblock Springer Nature, 2015.

\bibitem{Ref31a}
E.L. Saldin, E.A. Schneidmiller, and M.V. Yurkov.
\newblock Radiative interaction of electrons in a bunch moving in an undulator.
\newblock {\em Nuclear Instruments and Methods in Physics Research Section A:
  Accelerators, Spectrometers, Detectors and Associated Equipment},
  417(1):158--168, 1998.

\bibitem{Ref31b}
Juhao Wu, Tor~O. Raubenheimer, and Gennady~V. Stupakov.
\newblock Calculation of the coherent synchrotron radiation impedance from a
  wiggler.
\newblock {\em Physical Review Special Topics-Accelerators and Beams},
  6(4):040701, 2003.

\bibitem{Ref32}
R~Bonifacio, C~Pellegrini, and LM~Narducci.
\newblock Collective instabilities and high-gain regime free electron laser.
\newblock In {\em AIP conference proceedings}, volume 118, pages 236--259.
  American Institute of Physics, 1984.

\bibitem{Ref34}
Albert Hofmann.
\newblock {\em The physics of synchrotron radiation}, volume~20.
\newblock Cambridge University Press, 2004.

\bibitem{Ref36}
X.J. Deng.
\newblock private communication.

\bibitem{Ref37}
Y.~Zhang, X.J. Deng, Z.L. Pan, Z.Z. Li, K.S. Zhou, W.H. Huang, R.K. Li, C.X.
  Tang, and A.W. Chao.
\newblock Ultralow longitudinal emittance storage rings.
\newblock {\em Physical Review Accelerators and Beams}, 24(9):090701, 2021.

\bibitem{Ref38}
Sabor Elaydi.
\newblock {\em An Introduction to Difference Equations, 3rd. ed.}
\newblock Springer-Verlag New York, 2005.

\bibitem{Ref39}
Cheng-Ying Tsai, Juhao Wu, Chuan Yang, Moohyun Yoon, and Guanqun Zhou.
\newblock Sideband instability analysis based on a one-dimensional high-gain
  free electron laser model.
\newblock {\em Physical Review Accelerators and Beams}, 20(12):120702, 2017.

\bibitem{Ref40}
Cheng-Ying Tsai, Juhao Wu, Chuan Yang, Moohyun Yoon, and Guanqun Zhou.
\newblock Single-pass high-gain tapered free-electron laser with transverse
  diffraction in the postsaturation regime.
\newblock {\em Physical Review Accelerators and Beams}, 21(6):060702, 2018.

\bibitem{Ref41}
C.-Y. Tsai, C.~Emma, J.~Wu, M.~Yoon, X.~Wang, C.~Yang, and G.~Zhou.
\newblock Area-preserving scheme for efficiency enhancement in single-pass
  tapered free electron lasers.
\newblock {\em Nuclear Instruments and Methods in Physics Research Section A:
  Accelerators, Spectrometers, Detectors and Associated Equipment},
  913:107--119, 2019.

\bibitem{Ref42}
X.J. Deng, A.W. Chao, W.H. Huang, and C.X. Tang.
\newblock Courant-{S}nyder formalism of longitudinal dynamics.
\newblock {\em Physical Review Accelerators and Beams}, 24(9):094001, 2021.

\bibitem{Ref43}
Daniel Zwillinger and Alan Jeffrey.
\newblock {\em Table of integrals, series, and products}.
\newblock Elsevier, 2007.

\bibitem{Ref44}
G. Stupakov and S. Krinsky.
\newblock Derivation of {FEL} gain using wakefield approach.
\newblock {\em Proceedings of the 2003 Particle Accelerator Conference} (RPPG025).


\end{thebibliography}

\end{document}